\documentclass{aa}  

\usepackage{graphicx}
\usepackage{mathtools}
\usepackage{graphicx}	
\usepackage{gensymb}
\usepackage{multirow}
     
\usepackage{mathtools}

\usepackage{amssymb}	

\usepackage{amsmath}	
\usepackage{hyperref}%
\hypersetup{colorlinks=true,linkcolor=[rgb]{1.,0.2,0.2},citecolor=[rgb]{0.1,0.4,1.},filecolor=[rgb]{0.7,0.2,0.2},urlcolor=[rgb]{0.7,0.2,0.2}}
\newcommand{\massi}[1]{{\color{black}{#1}}}
\newcommand{\Extend}[1]{\mbox{{\sc refExtendedness~}}}

\usepackage{txfonts}
\begin{document}

   \title{Star–Galaxy Classification in Deep LSST Data with Random Forest:}

   \subtitle{A Pilot study on the Data Preview 1 Release}

   \titlerunning{Star–Galaxy Classification in Deep LSST Data with Random Forest}
   
   \author{M. Gatto
          \inst{1}\fnmsep\thanks{massimiliano.gatto@inaf.it}
          \and
          V. Ripepi
          \inst{1}
          \and
          M. Bellazzini
          \inst{2}
          \and 
          C. Tortora
          \inst{1}
          \and
          M. Dall'Ora
          \inst{1}
          }

   \institute{INAF-Osservatorio Astronomico di Capodimonte, Salita Moiariello, 16, 80131 Napoli, Italy
   \and
   INAF-Osservatorio di Astrofisica e Scienza dello Spazio, Via Gobetti, 93/3, 40129 Bologna, Italy
             }

   \date{}

 
  \abstract
   {The Vera C. Rubin Observatory Legacy Survey of Space and Time (LSST) will produce unprecedentedly deep and wide photometric catalogs, enabling transformative studies of faint stellar systems such as the research of ultra-faint dwarf galaxies (UFDs). A critical challenge for these studies is reliable star–galaxy separation at faint magnitudes, where compact background galaxies increasingly contaminate stellar samples.} 
   {This work aims to assess the performance of supervised machine-learning techniques for star–galaxy separation in LSST-like data, quantify the relative importance of morphological and photometric information, and identify the most effective combinations of input features for minimizing galaxy contamination while preserving stellar completeness in the faint regime relevant for UFD searches.}
   {We apply a Random Forest classifier to observations of the Extended Chandra Deep Field South from LSST Data Preview 1 (DP1), the deepest field observed within the DP1. We construct a curated sample of bona fide stars and galaxies using spectroscopic data, Gaia DR3, and multi-band photometric catalogs. We train and validate the classifier using several configurations of LSST-based input features, including multi-band colors, the LSST morphological parameter {\sc refExtendedness}, and photometric uncertainties.}
   {We find that LSST multi-band photometry alone delivers a good star–galaxy separation, significantly outperforming morphology-based classification at faint magnitudes. 
   Colors involving the $u$ band are essential to provide a robust star galaxy separation. Furthermore, explicitly including photometric uncertainties as input features yields the best overall performance. Across all configurations that include all the six LSST filters, galaxy contamination remains negligible almost the whole magnitude range probed in this work (i.e. $r \lesssim$ 27.5 mag).}
   {Our results demonstrate that supervised machine-learning methods, when combined with LSST multi-band photometry, can effectively suppress galaxy contamination in deep stellar catalogs, ensuring that searches for UFDs are not significantly compromised. Given that the DP1 data are shallower and have poorer seeing than the final LSST survey, our findings should be regarded as a conservative lower limit on the performance achievable with the full 10-year dataset. To facilitate further development, we will publicly release the curated star–galaxy sample used in this work.}

   \keywords{Methods: data analysis, Methods: statistical, Techniques: photometric, Surveys, Catalogs
               }

   \maketitle
%

\section{Introduction}
\label{sec:intro}

Since the beginning of the new millennium, our understanding of the formation and evolution of the Milky Way (MW) has advanced substantially, largely owing to wide and deep panchromatic surveys conducted with both ground-based facilities, such as the Sloan Digital Sky Survey (SDSS\footnote{https://www.sdss.org/}), the Panoramic Survey Telescope and Rapid Response
System \citep[Pan-STARRS;][]{Chambers-2016}, the Dark Energy Survey\footnote{https://www.darkenergysurvey.org/} (DES), and space missions such as Gaia \citep{Gaia-Prusti-2016}. 
These surveys have produced extensive and accurate stellar catalogs, enabling detailed reconstructions of the MW’s past and ongoing mass assembly, and, more broadly, of the evolution of the Local Group.
Most current wide-field surveys reach an average depth of $r \simeq 24$ mag, meaning they are able to resolve main-sequence turn-off (MSTO) stars belonging to an old stellar population ($t \sim 12$ Gyr) at a heliocentric distance of $\sim 150$ kpc.\par
The Vera C. Rubin Observatory\footnote{https://rubinobservatory.org/} (VRO), now in operation, will dramatically extend this capability through the 10-year Legacy Survey of Space and Time (LSST). The LSST will repeatedly image the Southern Hemisphere in six photometric bands ($ugrizy$), ultimately achieving a coadded depth of $r \simeq 27.5$ mag \citep{Ivezic-2019}. At this depth, main-sequence stars will be resolved well beyond the presumed edge of the Galactic halo. An immediate science case enabled with this unprecedented vast, deep and precise catalogue of stars will be the detection of dwarf galaxies, by means of their resolved stellar population, out to $\sim $5 Mpc across the entire Southern sky \citep{Mutlu-Pakdil-2021}.\par
However, realizing this potential requires an effective and reliable star–galaxy classifier, particularly at faint magnitudes where contamination from unresolved galaxies becomes severe. \citet{Fadely-2012} demonstrated that, in the color range typical of old, metal-poor MSTO stars inhabiting the MW halo ($g-r \lesssim 1.0$ mag), unresolved galaxies begin to outnumber stars at $r \gtrsim 23.5$ mag, with their relative fraction increasing sharply toward fainter magnitudes. 
This contamination poses a major obstacle for the search of Ultra-Faint Dwarf (UFD) galaxies, the most ancient, metal-poor, and dark matter-dominated galaxies known to date \citep{Simon2019}. 
Detection techniques for UFDs rely on identifying overdensities of old, metal-poor stars over limited regions of the sky \citep[e.g.,][]{Walsh-2009}, and inadequate star–galaxy separation can therefore lead to catastrophic misclassifications, such as mistaking a projected overdensity of unresolved background galaxies (e.g., a distant group or cluster) for a nearby dwarf galaxy.\par
In the absence of spectroscopic radial velocity measurements, whose acquisition over large portion of the sky is prohibitively expensive in terms of telescope time, source classification generally relies on morphological diagnostics. Standard pipelines exploit shape parameters, such as source roundness, or flux ratio tests.  
In the LSST pipeline, for example, the {\sc Object} catalog\footnote{The {\sc Object} catalog provides measurements of objects detected in deep coadd images, with a signal-to-noise ratio (SNR) of $>$5 in at least one band, \citep{NSF-DOE-VRO-2025a}.} includes the {\sc refExtendedness} parameter, which serves as a binary star–galaxy discriminator \citep{VRO-DP1}. This parameter is assigned based on the ratio between cModel\footnote{\massi{cModel is a composite flux obtained from the linear combination of de Vaucouleurs and exponential fluxes in each band (see https://dp1.lsst.io/tutorials/notebook/201/notebook-201-1.html).}} and PSF fluxes, with sources classified as point-like if $f_{\rm psf} \geq 0.985  f_{\rm cmodel}$\footnote{https://sdm-schemas.lsst.io/dp1.html} \citep{Choi-2025}. Nonetheless, the classification degrades at faint magnitudes: tests with artificial source injection in fields observed in the LSST Data Preview 1 (DP1) indicate that the fraction of correctly classified stars drops to $\sim 50\%$ at $i = 23.8$ mag, and at $i \simeq 24.5 - 25$ mag nearly 20\% of all sources are incorrectly classified \citep[see][their Section 5.6]{VRO-DP1}.\par
To improve upon such limitations, several studies have explored more sophisticated approaches, often leveraging multi-band photometry. \citet{Fadely-2012} compared maximum likelihood methods, hierarchical Bayesian classifiers, and support vector machines (SVMs), applying them to the 30-band COSMOS survey \citep{Scoville-2007}, which reaches $r \simeq 25$ mag over 2 deg$^2$. They found SVMs, and hierarchical Bayesian classifiers, yield the most effective star–galaxy separation.
\citet{Logan&Fotopoulou2020} applied the Hierarchical Density-Based Spatial Clustering of Applications with Noise ({\sc HDBSCAN}) to a dataset of about 50,000 spectroscopically labeled sources from the sample of \citet{Fotopoulou&Paltani2018}, aiming to distinguish stars, galaxies, and quasars in a multidimensional color space ($ugrizYHJKW_1W_2$). They achieved an accuracy of $\geq$ 98\% down to $i \simeq 24-25$ mag.
\citet{Jeakel-2026} applied an XGBoost classifier trained on the $\sim$60-band Javalambre Physics of the Accelerating Universe Astrophysical Survey\footnote{https://www.j-pas.org} (J-PAS), combining photometric and morphological features to distinguish stars and galaxies down to $r \sim 23.5$ mag, showing that incorporating morphology significantly enhances performance.
\citet{Zhang-2025} combined a Bayesian photometric redshift estimator with a multilayer perceptron to separate artificially injected stars and galaxies in Kilo-Degree Survey (KiDS) Data Release 4–like images \citep{Kuijken-2019}. They reported a negligible false positive rate (FPR) down to ${\rm MAG\_AUTO} \leq 25$ mag, although the true positive rate (TPR) drops below 40\% at the same magnitude.
\massi{\citet{Khramtsov-2019} applied CatBoost, a gradient-boosting decision-tree algorithm, to the nine-band ($ugriZYJHK_s$) KiDS DR4 + VIKING dataset, obtaining TPRs of $\simeq 98$–99\% down to $r \sim 22$ mag for stars, galaxies, and quasars.}
More recently, \citet{Feng-2025} employed a multimodal neural network on the KiDS Data Release 5 \citep{Wright-2024} + VIKING nine bands survey, to classify stars, galaxies, and quasars, achieving an overall accuracy of $\sim 99\%$ down to $r \sim 23$ mag.
Nevertheless, these efforts probe significantly shallower regimes than LSST’s final depth ($r \simeq 27.5$ mag), and thus may not fully capture the classification challenges LSST will face.\par
The purpose of this work is to show that a straightforward, photometry-based machine-learning model can already suppress much of the unresolved galaxy contamination that would otherwise hinder UFD searches in LSST, even though future, more advanced techniques may further improve upon the method presented here.
To this aim, we test the application of a random forest classifier, a supervised machine learning algorithm, for star–galaxy separation in the recent released LSST Data Preview 1 \citep[DP1;][]{VRO-DP1}.
Throughout this work, we explore multiple configurations of input features, primarily constructed from LSST multi-band photometry, and optionally incorporating additional features (including, but not limited to, the \Extend~parameter provided in the DP1 catalog), as well as cases in which selected features are intentionally excluded to assess their impact on classification performance.
Color–color diagrams are known to be powerful for this task, as stars occupy relatively narrow loci while galaxies are more broadly distributed, thereby enhancing separability. In particular, optical–infrared color combinations have proven to be robust discriminants \citep[e.g.,][]{Maddox-2008, Tortora-2018}.
To emulate LSST’s final depth, we exploit the Extended Chandra Deep Field-South (ECDFS) region available within LSST DP1, which has been observed multiple times. \massi{Although this field represents the deepest dataset currently released by LSST, its properties remain significantly shallower than the expected 10-year survey. In particular, the ECDFS area reaches a 5$\sigma$ point-source depth of only $r \simeq 26$ mag \citep{VRO-DP1}, about 1.5 mag brighter than LSST’s nominal coadded depth. Moreover, the median point-spread function FWHM is $\simeq$ 1.14\arcsec \citep{VRO-DP1}, considerably worse than the $\simeq$ 0.65\arcsec median seeing anticipated for LSST operations \citep{Ivezic-2019}. As a consequence, both the depth and image quality of DP1 fall short of what the final survey will deliver.
These limitations imply that the performance we obtain in this work should be regarded as a conservative lower bound on what a random forest classifier may achieve when applied to the full 10-year LSST dataset.}\par
The paper is structured as follows: Sect. \ref{sec:data} presents the LSST DP1 dataset and the literature catalogs used to assemble the sample of bona-fide stars and galaxies. Section \ref{sec:rf} details the optimization of the random-forest hyperparameters and discusses the resulting classification performance for different choices of input features. Finally, Sect.~\ref{sec:discussion} provides a summary and outlines the implications of this work.

\section{Data}
\label{sec:data}

\subsection{The LSST Data Preview 1}

The Data Preview 1 (DP1) represents the first data release obtained with the Vera C. Rubin Observatory \citep{VRO-DP1}. DP1 includes raw and calibrated single-epoch images and their corresponding {\sc Source} catalogs, as well as stacked deep coadd images and the associated {\sc Object} catalogs, among other products. A detailed description of the DP1 data products is provided in the official documentation \citep{SF-DOE-VRO-2025b}.
DP1 is based on 48 nights of observations collected during the first on-sky commissioning campaign with the LSSTComCam. The dataset covers a total area of $\sim 15$ deg$^2$, distributed across seven distinct, non-contiguous fields. Details of the LSSTComCam campaign and of DP1 can be found in SITCOM-149\footnote{https://sitcomtn-149.lsst.io/} and in the DP1 overview paper \citep{VRO-DP1}.\par
For the purposes of this work, we focus on the Extended Chandra Deep Field-South (ECDFS) field, which benefits from the most extensive and homogeneous temporal coverage among the DP1 fields. In particular, ECDFS was observed 43, 230, 237, 162, 153, and 30 times in the $u$, $g$, $r$, $i$, $z$, and $y$ filters, respectively. This cadence enables a \massi{5$\sigma$ point-source depth of $r \simeq 26$ mag.}
We retrieved the data from the {\sc Object} table, which contains measurements of sources detected in the coadded images. The ADQL query used for the extraction is the following:
\begin{verbatim}
SELECT objectId, coord_ra, coord_dec, ebv,  
refExtendedness, u_psfMag, u_psfMagErr, 
u_psfFlux_flag, g_psfMag, g_psfMagErr, 
g_psfFlux_flag, r_psfMag, r_psfMagErr, 
r_psfFlux_flag, i_psfMag, i_psfMagErr, 
i_psfFlux_flag, z_psfMag, z_psfMagErr, 
z_psfFlux_flag, y_psfMag, y_psfMagErr, 
y_psfFlux_flag
FROM dp1.Object 
WHERE CONTAINS(POINT('ICRS',coord_ra,coord_dec), 
CIRCLE('ICRS', 53.16, -28.1, 1.0)) = 1
\end{verbatim}
We retained only sources for which the $\_psfFlux\_flag$ was not set, as a triggered flag indicates that the PSF photometry in that band failed and was therefore replaced by forced photometry.
In addition, we imposed a bright-end cut to remove saturated sources.
Specifically, \citet{Choi-2025} estimated the saturation bright limit to be $\sim$ 15.70, 15.08, 15.19 mag in the $g$, $r$, and $i$ bands, respectively. We adopted these values as upper limits on magnitudes. 
This query returned 356,786 objects within a circular region of radius $40~\arcmin$ (area $\sim 1.4$ deg$^2$), centered at (RA, Dec) = (53.16\degree, -28.1\degree).

\subsection{Spectroscopic catalogs}
\label{sec:spectro_cat}

\begin{table}[]
    \caption{Number of stars and galaxies retrieved from external catalogs.}
    \tiny
    \hspace{-1.cm}
    \begin{tabular}{l|c|c|l}
    \hline
      Catalog & N$_{\rm stars}$ & N$_{\rm galaxies}$ & Reference\\
      \hline\hline
       3D-HST & 374 & 3920 & \citet{Momcheva-2016} \\
       GOODS/VIMOS & 50 & 1105 & \citet{Popesso-2009, Balestra-2010}\\
       GOODS/FORS2 & 0 & 130 & \citet{Vanzella-2008}\\
       GMASS & 0 & 22 & \citet{Kurk-2013}\\
       CANDELS & 0 & 213 & \citet{Kodra-2023}\\
       VANDELS & 1 & 455 & \citet{Talia-2023}\\
       MUSE HUDF & 0 & 381 & \citet{Bacon-2023}\\
       JADES & 0 & 194 & \citet{D'Eugenio-2025}\\
       VVDS & 0 & 634 & \citet{LeFevre-2013}\\
       MUSYC & 39 & 143 & \citet{Cardamone-2010}\\
       OzDES & 146 & 972 & \citet{Lidman-2020}\\
       ACES & 125 & 3081 & \citet{Cooper-2012} \\
       GAIA DR3 & 2004 & 0 & \citet{Gaia-Vallenari-2023}\\
       SIMPLE & 165 & 0 & \citet{Damen-2011}\\
       COMBO-17 & 178 & 0 & \citet{Wolf-2004}\\
       DESI & 1279 & 0 & \citet{Duncan-2022}\\      
         \hline
         \hline
         TOTAL & 4,361 & 11,250 & \\
         \hline
    \end{tabular}
    
    \label{tab:lit_catalogs}
\end{table}

Random Forest is a supervised machine learning algorithm, meaning that it learns to classify sources based on a training set of objects with secure labels, i.e. sources confidently identified as stars or galaxies. 
To construct a robust training set, we cross-matched our LSST catalog with spectroscopic surveys available in the literature that overlap with the ECDFS region. Spectroscopic redshifts provide an unambiguous physical discriminator between stars and galaxies, making them an ideal source of ground truth labels.
Specifically, we used the following catalogs:\par
    The 3D-HST survey is a 248-orbit Treasury program with the Hubble Space Telescope (HST) that employed the WFC3/G141 slitless grism to obtain near-infrared spectra of galaxies up to redshift $z < 3$ \citep{Momcheva-2016}. The survey covers five deep extragalactic fields, including one centered on the Chandra Deep Field South (CDFS), and is complemented by extensive multiwavelength imaging from both ground- and space-based facilities, enabling high-quality photometric redshift estimates \citep[see also][]{Skelton-2014}. 
    The spectroscopic catalog contains 4,795 spectroscopic redshifts, with stars identified by a value of $z_{\rm best} = -1$.\footnote{The full catalog also includes photometric redshifts for more than 45,000 sources. We opted to restrict only to sources with spectroscopic measurements.} 
    Cross-matching with ECDFS (within a tolerance of 1\arcsec) yielded 4,294 
    objects, the majority being galaxies (see Table~\ref{tab:lit_catalogs}).\par   
    The Great Observatories Origins Deep Survey (GOODS) targeted two deep fields: Hubble Deep Field North (GOODS-N) and CDFS (GOODS-S), with multiwavelength coverage from infrared to X-ray, aiming to address fundamental questions of galaxy and AGN formation, as well as the distribution of baryonic and dark matter at high redshift \citep{Vanzella-2008, Popesso-2009, Balestra-2010}.
    Spectroscopic data were obtained with VIMOS at VLT, yielding a total of 3,218 redshifts, with stars classified as Star in the column COMM. We selected 2,223 sources with high-quality flags (QF = A or B), of which 
    1,155 matched counterparts in our ECDFS catalog, again predominantly galaxies.\par
    As part of the GOODS project, \citet{Vanzella-2008} carried out deep spectroscopic observations of GOODS-S with the FORS2 at VLT instrument. Of the 1,165 available redshifts, (\massi{with stars classified as Star in the column comments}), we selected 972 with quality column = A or B. The cross-match added 130 sources to our secure catalog, all of them are galaxies.
    The Galaxy Mass Assembly ultra-deep Spectroscopic Survey (GMASS) was designed to probe both massive quiescent and star-forming galaxies at $z > 1.4$, with particular focus on the peak epoch of galaxy assembly around $z \sim 2$ \citep{Kurk-2013}. Targets were drawn from the CDFS/GOODS-S field. Of the 210 spectroscopic redshifts available, we selected 192 with reliable quality flags ($\rm {q\_zsp} = 1$). The crossmatch with ECDFS field yielded further 22 sources, all of which correspond to galaxies.\par
    The Cosmic Assembly Near-Infrared Deep Extragalactic Legacy Survey (CANDELS) combines high spatial resolution imaging from HST with complementary intermediate-resolution optical and IR imaging. 
    \citet{Kodra-2023} published the first large-scale release of photometric redshifts derived by the CANDELS collaboration, based on a spectroscopic training set of 5,807 high-quality redshifts covering all five survey fields. Within the GOODS-S region, 2,312 sources possess reliable spectroscopic redshifts. Our cross-match with the ECDFS catalog identified 213 unique counterparts, all of which are galaxies.\par
    The VANDELS survey is a deep spectroscopic campaign carried out with the VIMOS at VLT, designed to probe in detail the physical properties of high-redshift galaxies. It targeted approximately 2,100 galaxies within the redshift range $1 < z < 6.5$, located in the CDFS and Ultra Deep Survey (UDS) regions. \citet{Talia-2023} presented the final VANDELS data release, which includes 880 sources in the CDFS field, 679 of which have secure spectroscopic redshifts (column $\rm {zflg}$ = 3 or 4). Our cross-match with the ECDFS catalog yielded 456 common sources,all but one of which are galaxies; the only stellar object in the sample corresponds to the source with spectroscopic redshift equal to zero.\par
    The MUSE Hubble Ultra Deep Field (HUDF) survey represents one of the deepest spectroscopic explorations ever conducted of the GOODS-S field. The second public data release \citep{Bacon-2023} includes a total of 2,221 extracted spectra, of which 1,711 have secure redshift determinations (column $\rm zconf$ = 2 or 3). Our cross-match with the ECDFS catalog yielded 381 new galaxies.\par
    The JWST Advanced Deep Extragalactic Survey (JADES) is an observational program designed to particularly investigate galaxies at $z > 3$. \citet{D'Eugenio-2025} presented the third data release of JADES, which combines deep NIRCam imaging with extensive NIRSpec spectroscopy over the GOODS-N and GOODS-S fields. From a total of 2,525 extracted spectra, 914 have high-quality redshift flags ($\rm z\_Spec\_flag$ = a or b). Our cross-match with the ECDFS catalog resulted in 
    194 associations, all corresponding to galaxies.\par
    The VIMOS VLT Deep Survey (VVDS) represents one of the most extensive spectroscopic efforts to map galaxy evolution across cosmic time, since $z \sim 6.7$. Spectroscopic observations were carried out with the VIMOS at VLT. The final release includes more than 45,000 sources \citep{LeFevre-2013}. Among these, 1,323 sources in the ECDFS have reliable spectroscopic redshifts ($f_z > 1$). Our cross-match yielded 634 counterparts, all identified as galaxies.\par
    The Multiwavelength Survey by Yale–Chile (MUSYC) provides deep optical and near infrared imaging across four different fields. In particular, \citet{Cardamone-2010} presented the dataset covering the ECDFS region, obtained with the Subaru Telescope using 18 medium-band filters. These observations were complemented with broad-band optical and near-infrared data ($UBVRIzJHK$) as well as Spitzer/IRAC imaging, resulting in a uniform multiwavelength catalog. To assess the accuracy of the derived photometric redshifts, they adopted spectroscopic measurements available from the literature. From the spectroscopic subsample of 
    257 high confident redshifts ($\rm q\_zsp \geq2$), our cross-match with the ECDFS catalog yielded 
    182 sources: 
    39 stars (sources with spectroscopic redshift equal to zero, and 143 galaxies.\par
    The Australian Dark Energy Survey (OzDES) is a spectroscopic campaign conducted with the 2dF fibre positioner and the AAOmega spectrograph on the 3.9-m Anglo-Australian Telescope. Its primary goal was to complement the Dark Energy Survey (DES) by obtaining spectroscopic redshifts for transient hosts, active galactic nuclei, and galaxies located within the ten DES deep fields. \citet{Lidman-2020} presented the second OzDES data release, comprising nearly 30,000 reliable redshifts ($q_{\rm op}$ = 3, 4 for galaxies and $q_{\rm op}$ = 6 for stars) from a total of 38,624 spectra. Our cross-match with the ECDFS catalog returned 
    1,118 unique sources, including 146 stars, and 972 galaxies.
    The Arizona CDFS Environment Survey (ACES) is a spectroscopic program carried out with the IMACS spectrograph on the Magellan-Baade telescope, designed to significantly improve the sampling of galaxy redshifts across the ECDFS. From the full ACES catalog of 12,983 entries, 6,601 objects have high-quality redshift classifications \citep[stars with Q = -1, galaxies with q = 3, 4;][]{Cooper-2012}. Our cross-match with the ECDFS catalog identifies 3,206 matched sources, including 125 stars and the remainder classified as galaxies.\par
Table~\ref{tab:lit_catalogs} summarizes the number of stars and galaxies obtained from each spectroscopic catalog after cross-matching with our sample of sources.

\subsection{Gaia Data Release 3}

The cross-matched catalog from spectroscopic surveys yielded a total of 11,985 sources, the vast majority of which — 11,250 objects, or $\sim$94\% — are galaxies. Such an imbalanced training set is problematic, as it fails to adequately sample the photometric parameter space of both classes. 
To mitigate this imbalance and increase the number of securely identified stars, we cross-matched ECDFS with Gaia Data Release 3 \citep{Gaia-Vallenari-2023}, which provides precise parallaxes and proper motions that are highly effective for distinguishing Galactic stars from distant galaxies. The cross-match yielded 2,770 additional sources with no prior spectroscopic counterpart. Of these, 2,088 have reliable astrometric solutions with RUWE $<$ 1.4. Restricting further to sources with a star probability $P_{\rm SS} > 0.99$, we obtained a final subsample of 2,004 high-confidence stars.
The combined catalog therefore comprises 13,989 sources, of which 2,630 are stars (19\%) and 11,359 are galaxies (81\%).\par

\subsection{Multi-photometric catalogs}
\label{sec:multi-photometric catalogs}

Given the relatively bright limiting magnitude of Gaia, the resulting stellar subsample is dominated by sources with $r \leq 21$ mag. A potential concern in relying solely on these objects is that a machine-learning classifier trained only on bright stars may not generalize well to the much fainter regime, where the separation between stars and unresolved galaxies becomes significantly more challenging.
To mitigate this limitation, we supplemented the stellar sample with stars identified through multi-band photometry, leveraging catalogs that include either a large number of photometric bands or mid-infrared observations, where stellar and galaxy spectral energy distributions diverge more strongly.
It is worth noting that although some of these stars were originally classified based on high-dimensional photometric information, the goal of this study is to assess how effectively a machine-learning classifier can separate stars and galaxies, \massi{and can reduce the contamination from compact galaxies at faint magnitudes,} using only LSST filters.
In this sense, our approach should be regarded as a pilot investigation aimed at testing the viability and performance of such methods in the photometric regime relevant for LSST. The inclusion of stars identified from deep multi-band data simply allows the classifier to sample the regions of color–color space where Gaia becomes incomplete, thus ensuring that the model is trained across the full dynamic range of interest.\par 
We adopted the following catalogs to increase the number of stars at faint magnitudes:
\begin{itemize}
    \item The SIMPLE survey (Spitzer IRAC/MUSYC Public Legacy Survey in the Extended CDF-S) provides deep IRAC imaging (3.6, 4.5, 5.8, and 8.0 $\mu$m) over $\simeq$1600 arcmin$^2$ surrounding the GOODS-S field. The IRAC data are matched to MUSYC optical and near-infrared imaging ($UBVRI_zJHK$), yielding a catalog of 61,233 total sources, of which 43,782 are detected at $S/N > 5$ at 3.6 $\mu$m, with 19,993 objects having full 13-band photometry \citep{Damen-2011}. Our cross-match with the ECDFS catalog resulted in 
    165 stars.
    \item The COMBO-17 survey provides 17-band optical photometry (350–930 nm) covering the CDFS region. The catalog contains 63,501 sources with a multi-color classification into stars, galaxies, and QSOs \citep{Wolf-2004}. The cross-match with this catalog provided 178 stars.
    \item We also included stellar sources identified in the Dark Energy Spectroscopic Instrument (DESI) Legacy Imaging Surveys. The dataset combines optical ($grz$) and mid-infrared photometry (3.4, 4.6, 12, and 22 $\mu$m) from WISE and NEOWISE \citep{Duncan-2022}. From this catalog, we selected sources with a stellar probability $p\_star \geq 0.9$, providing further 
    1,279 likely stars to our catalog. 
\end{itemize}
{To assess the purity of the multi-photometric catalogs, we cross-matched them with spectroscopic catalogs described in Sect.~\ref{sec:spectro_cat}. A matching radius of 1\arcsec~was adopted. We obtained 2917 unique sources classified as stars in the multi-photometric catalogs and having a spectroscopic counterpart. Among them, 2488 are spectroscopically confirmed as stars, corresponding to a stellar purity of 85.3\%.}

\subsection{The full sample of bona-fide stars and galaxies}

\begin{figure}
    \centering
    \includegraphics[width=1.0\linewidth]{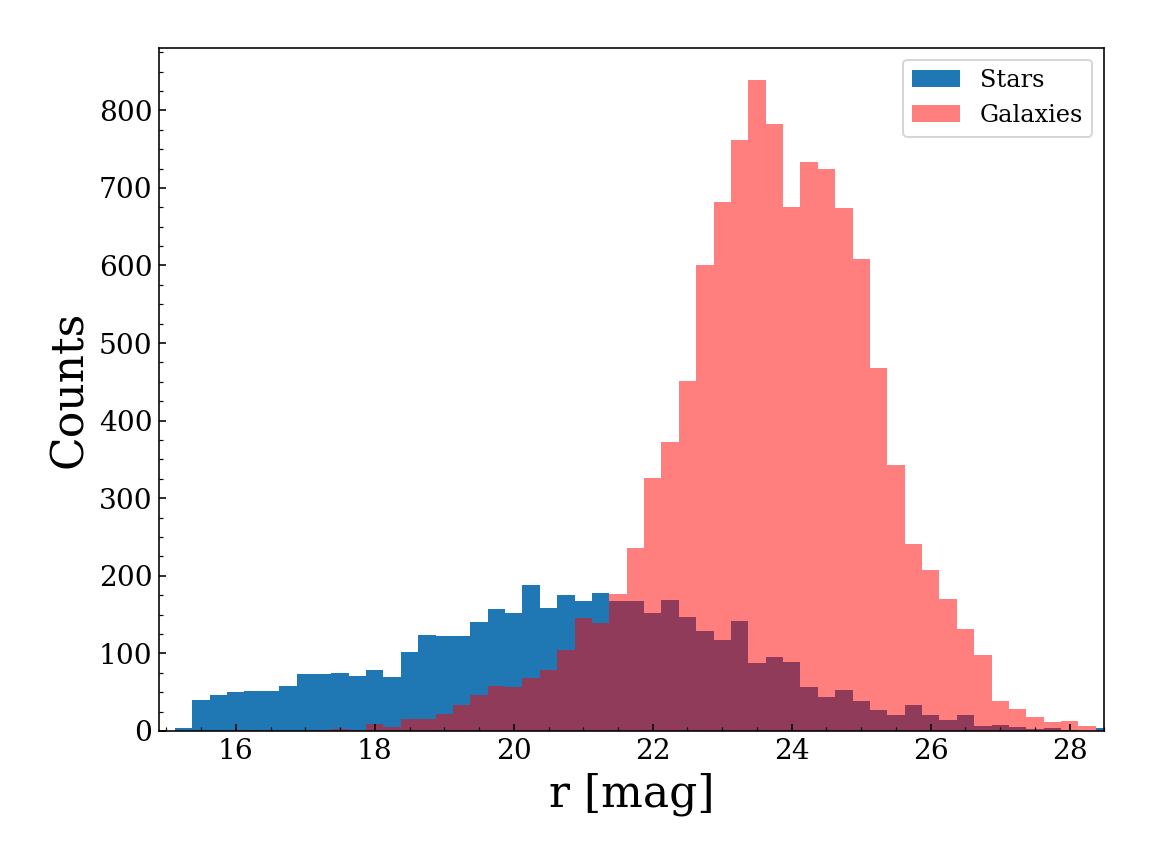}
    \caption{$r$-band magnitude distribution of the full sample of bona-fide stars and galaxies. Stars are shown in blue and galaxies in red.}
    \label{fig:histo_sample}
\end{figure}

\begin{figure}
    \centering
    \includegraphics[width=1.0\linewidth]{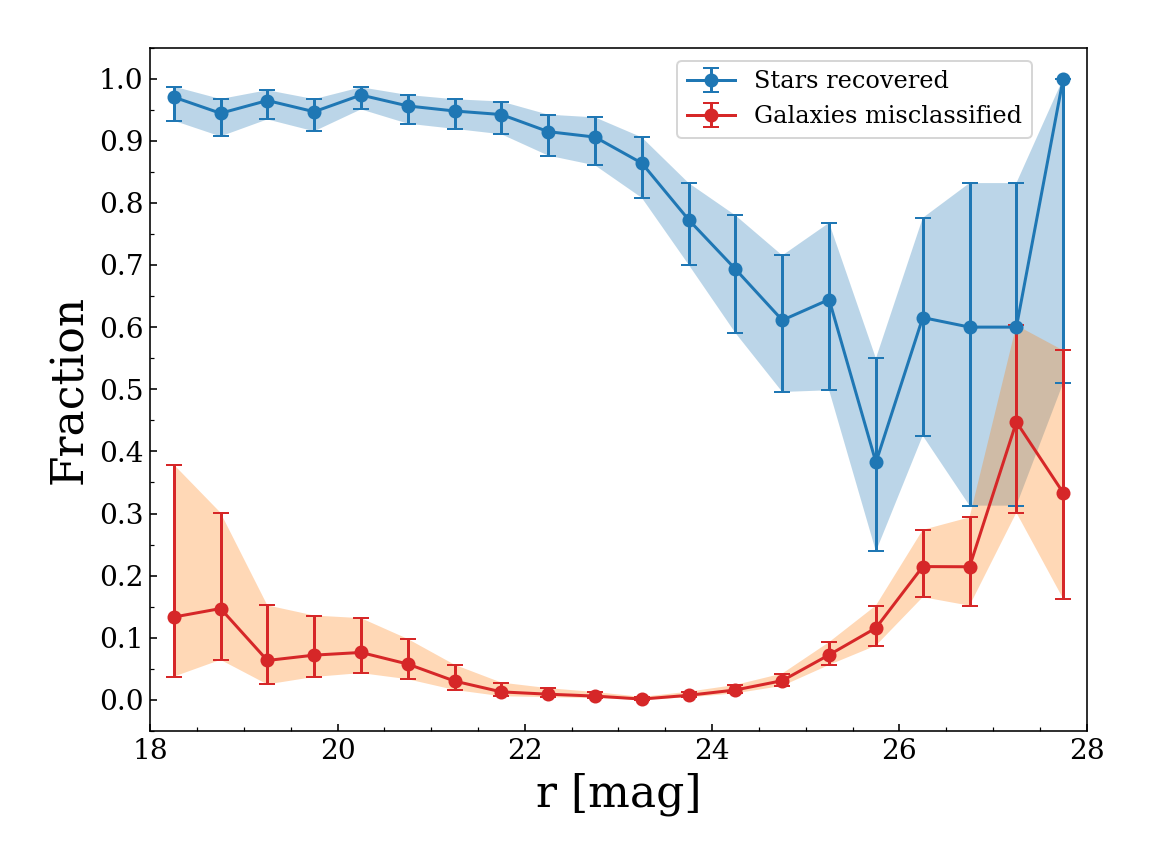}
    \caption{Performance of the {\sc refExtendedness} parameter as a function of $r$-band magnitude for our catalog. The blue solid line shows the fraction of true stars correctly identified as stars, while the red solid line indicates the fraction of true galaxies misclassified as stars.}
    \label{fig:class_fraction_lsst}
\end{figure}

Our final curated catalog contains 15,611 sources, of which 4,361 are classified as stars (28\%) and 11,250 as galaxies (72\%), as summarized in Table~\ref{tab:lit_catalogs}.
\massi{This final catalog will be made publicly available for future scientific use.}
Figure~\ref{fig:histo_sample} displays the $r$-band magnitude distribution of this final sample. Stars (blue histogram) predominantly occupy the bright regime, with a peak at $r \simeq 21$ mag and a rapid decline beyond $r \approx 24$ mag. Conversely, galaxies (red histogram) peak at significantly fainter magnitudes ($r \sim 23 - 24$ mag). 
This behavior is naturally linked to the characteristics of the input catalogs: many of the stars in our sample originate from Gaia DR3 or from high–quality multi-band surveys, both of which preferentially include relatively bright stellar sources. 
At the same time, this distribution is also representative of what is expected observationally: samples tend to be star-dominated at bright magnitudes, while progressively fainter magnitudes are increasingly populated by compact galaxies \citep{Fadely-2012}. 
The decline in the number of stars at $r \gtrsim 24$ mag in our compiled catalog therefore mirrors the observational regime in which searches for UFDs are typically conducted.
Importantly, this does not hinder the goals of the present work. Our aim is to assess whether machine-learning classifiers trained on LSST colors can effectively reduce the contamination from unresolved galaxies in stellar samples. This aspect is particularly critical for detecting very low-surface-brightness systems in the Local Volume, where contamination, rather than completeness, might be the limiting factor.\par
Figure~\ref{fig:class_fraction_lsst} illustrates the performance of the {\sc refExtendedness} morphological classifier on our sample. \massi{In this case, we plotted only the 14,360 sources that have complete photometry in all six LSST filters and a valid value of the {\sc refExtendedness} parameter.} The fraction of stars correctly identified decreases steadily with magnitude, reaching $\simeq$60\% at $r \simeq 24.5-25$ mag, and is subject to large uncertainties due to the small number of stars in this regime. In contrast, galaxy contamination \massi{steadily increases from $r \simeq 24.5$ mag} and remains below $\sim$20\% up to $r \sim 26$ mag before rising sharply, exceeding $\sim$40\% at $r \gtrsim 27$ mag.
This confirms that {\sc refExtendedness} is effective at bright and intermediate magnitudes but loses discriminating power near the LSST detection limit, where morphology alone becomes unreliable. 
Nevertheless, its relatively low galaxy contamination over a broad magnitude range persuaded us to \massi{test also the inclusion of} {\sc refExtendedness} as an additional feature in our random forest classifier.

\section{The Random Forest Classifier}
\label{sec:rf}

Random Forest (RF) is a supervised ensemble learning algorithm widely used for classification tasks \citep{randomforest}. It operates by constructing a large number of decision trees, each trained on a subset of the training data. At each split in a tree, only a random subset of features is considered, which introduces additional randomness and decorrelation among the trees, thus mitigating overfitting. 
RF classifiers have several properties that make them particularly well suited to star–galaxy separation. First, they can naturally handle heterogeneous and correlated input features (such as colors derived from LSST magnitudes), without requiring strong assumptions on the underlying data distribution. Second, they are robust against noisy measurements and outliers, a key aspect given the faint magnitude regime probed by LSST. Third, RFs provide a quantitative estimate of feature importance, allowing us to identify which color combinations contribute most effectively to the separation between stars and galaxies.\par 
\massi{To optimize model performance during photometric analysis, we de-reddened the stellar magnitudes across all six filters using the coefficients provided by \citet{Schlafly&Finkbeiner2011}.}
In the next sections, we describe the procedures we adopt to set RF parameters, train the model, and test its performance.

\subsection{Tuning the Random Forest parameters}
\label{sec:rf_tuning}

To implement the RF classifier we used the Python library {\it scikit-learn} \citep{scikit-learn}, which provides a flexible and efficient framework for model training, evaluation, and prediction.
The performance of a Random Forest depends on a set of hyperparameters, such as the number of trees in the ensemble, the maximum depth of each tree, and the minimum number of samples required to split a node, that are not learned during training but must be specified by the user.
A common strategy is to explore the hyperparameter space and identify the combination that yields the best-performing model.
To assess model performance, a suitable evaluation metric must be chosen, such as accuracy, F1 score, or recall. In this work, we adopted the F1 score as the primary metric, which is defined as:
\begin{equation}
    \rm F1 = 2 \times \frac{precision \times recall}{precision + recall}
\end{equation}
where
\begin{equation}
    \rm precision = \frac{TP}{TP + FP}
\end{equation}
is the fraction of predicted positives that are correct, and
\begin{equation}
    \rm recall = \frac{TP}{TP + FN}
\end{equation}
is the fraction of true positives that are successfully recovered.
We split the dataset into a training set (70\% of the sources), and an evaluation sample of the performance of the model (30\% of the sources).\par
We employed the {\sc GridSearchCV} tool available in scikit-learn, which systematically evaluates all possible hyperparameter combinations within a user-defined grid through an exhaustive search. Although computationally expensive, since the model is trained and validated for every parameter combination, this approach ensures that the global optimum within the explored grid is identified. For completeness, scikit-learn also provides RandomizedSearchCV, which instead samples random subsets of the parameter space, but we opted for the exhaustive GridSearchCV strategy to guarantee robustness.\par
The following hyperparameter grid was explored:
\begin{itemize}
    \item n\_estimators = 100, 300, 500, 1000;
    \item criterion = gini, entropy;
    \item max\_depth = None, 10, 50, 100;
    \item max\_features = sqrt, log2, None;
    \item min\_samples\_split = 2, 5, 10;
    \item min\_samples\_leaf = 1, 2, 4;
    \item bootstrap = True, False
\end{itemize}
This corresponds to a total of 1,728 distinct parameter combinations. We emphasize that the goal of this work is not to provide an exhaustive discussion of the role of each parameter, for which we refer the interested reader to the official RandomForestClassifier documentation.\footnote{https://scikit-learn.org/stable/modules/generated/
sklearn.ensemble.RandomForestClassifier.html}\par
To reliably estimate the predictive performance of the classifier and to prevent overfitting to a single train/test split, {\sc GridSearchCV} internally adopts the k-fold cross-validation technique. 
In this approach, the training set is randomly divided into k equally sized subsets (folds). At each iteration, one fold is retained for validation while the remaining k-1 folds are used for training. This process is repeated k times, ensuring that each fold serves as validation exactly once. The final performance metric is then computed as the average across all folds, thus providing a more robust and less biased estimate of the model’s generalization ability compared to a single split.
In this work, we adopted k = 5.\par
\massi{In the following sections, we present the optimized hyperparameters and the corresponding performance metrics for each of the input feature configurations explored in this study. This allows us to assess how different combinations of LSST features influence the classifier’s ability to separate stars from galaxies across the full magnitude range.}

\subsection{Random Forest performance on the independent validation sample}
\massi{In this section, we evaluate the performance of the Random Forest classifier across multiple configurations of input features. Additional experiments, including further feature combinations, are presented in Appendix~\ref{app:rf_other_config}.}
\begin{table}[]
\caption{Performance metrics for the RF classifier on the validation sample for different combination of features.}
    \tiny
    \begin{tabular}{l|l|c|c|c|c}
    \hline
    Features & Class & Precision & Recall & F1 score & N objects\\
     &  & \% & \% & \% & \\
      \hline\hline
      \multirow{2}{*}{Reference Set} & Galaxies & 97.8 & 99.1 & 98.5 & 3109\\
     & Stars & 97.6 & 94.3 & 95.9 & 1199\\
     
         \hline 
         \multicolumn{6}{c}{Accuracy = 97.8\%}\\
         \hline\hline
     \multirow{2}{*}{No refExtendness} & Galaxies & 96.8 & 98.5 & 97.7 & 3109\\
     & Stars & 96.0 & 91.7 & 93.8 & 1199\\
    \hline 
         \multicolumn{6}{c}{Accuracy = 96.6\%}\\
         \hline
         \multirow{2}{*}{Ref. Set + Phot. Err.} & Galaxies & 98.2 & 99.5 & 98.8 & 3109\\
     & Stars & 98.6 & 95.2 & 96.9 & 1199\\
    \hline 
         \multicolumn{6}{c}{Accuracy = 98.3\%}\\
         \hline
         \multirow{2}{*}{No $u$-band} & Galaxies & 97.1 & 98.9 & 97.9 & 3271\\
     & Stars & 96.8 & 92.5 & 94.6 & 1291\\
    \hline 
         \multicolumn{6}{c}{Accuracy = 97.0\%}\\
         \hline
         \multirow{2}{*}{No $y$-band} & Galaxies & 97.5 & 98.7 & 98.1 & 3202\\
     & Stars & 96.3 & 93.1 & 94.5 & 1182\\
    \hline 
         \multicolumn{6}{c}{Accuracy = 97.2\%}\\
         \hline
         \multirow{2}{*}{Phot. Err. - No refExt.} & Galaxies & 97.3 & 99.0 & 98.1 & 3109\\
     & Stars & 97.2 & 92.9 & 95.0 & 1199\\
    \hline 
         \multicolumn{6}{c}{Accuracy = 97.3\%}\\
         \hline\hline    
    \end{tabular}
    
    \label{tab:metrics}
\end{table}
\begin{figure}
    \centering
    \includegraphics[width=.49\linewidth]{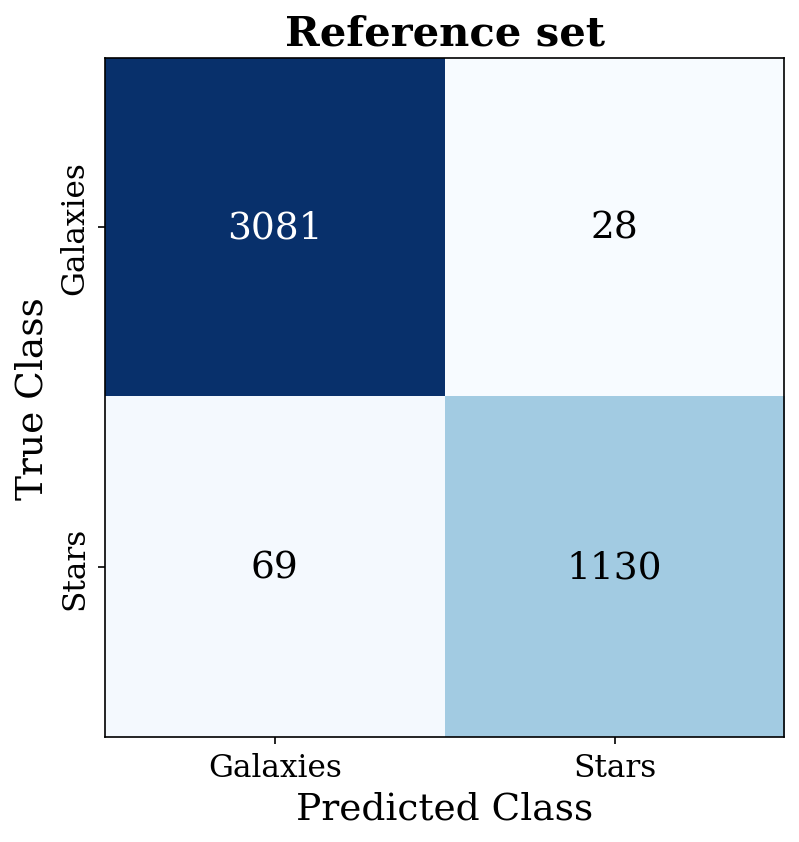}
    \includegraphics[width=.49\linewidth]{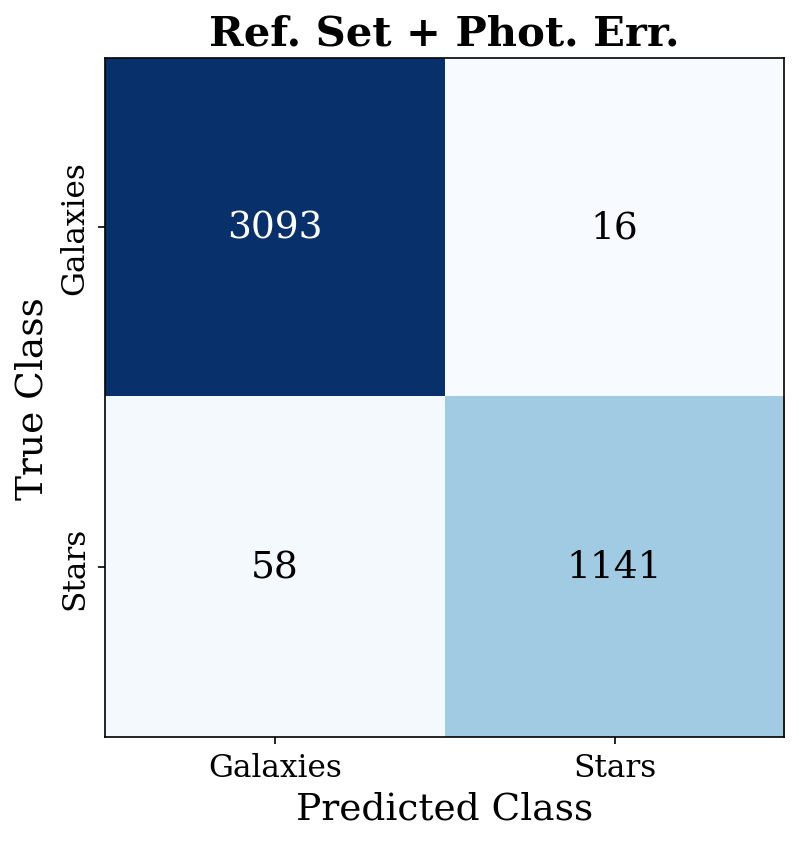}
    \caption{\massi{Confusion matrix on the validation sample for the reference set (left matrix) and with the inclusion of photometric uncertainties (right matrix).}}
    \label{fig:confusion_matrix}
\end{figure}
\begin{figure*}
    \centering
    \includegraphics[width=.24\linewidth]{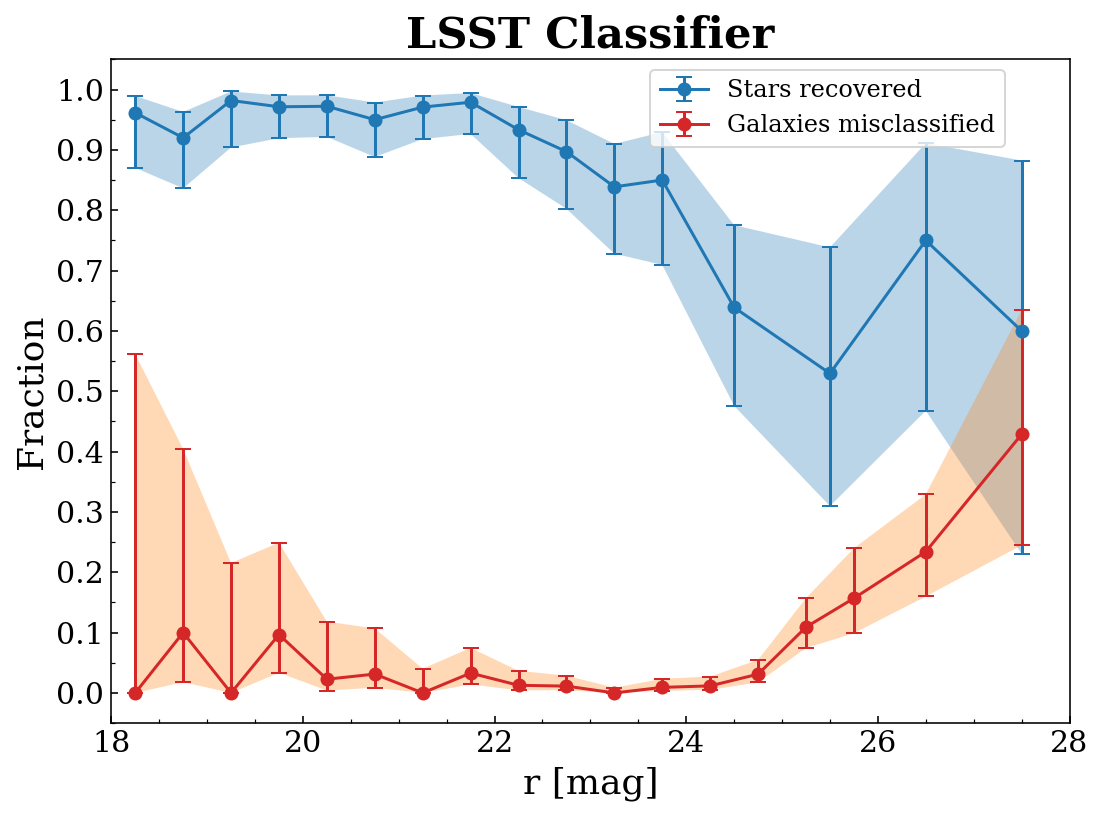}
    \includegraphics[width=.24\linewidth]{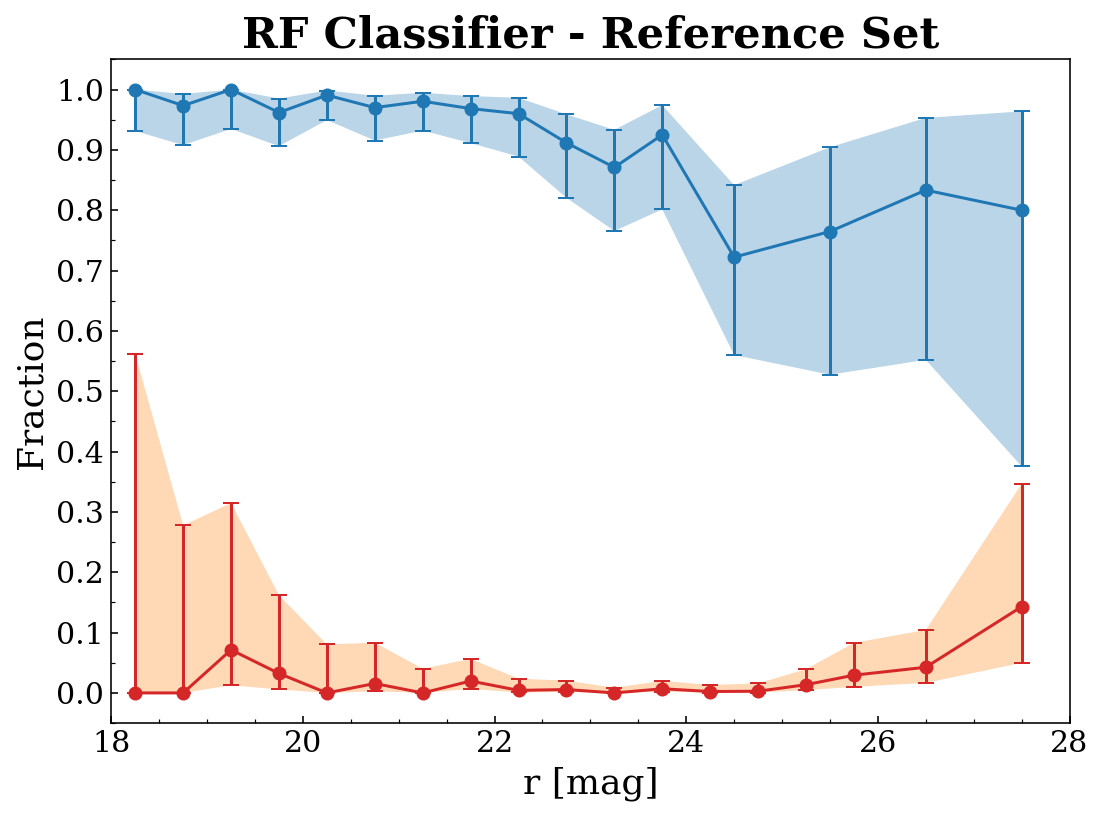}
    \includegraphics[width=.24\linewidth]{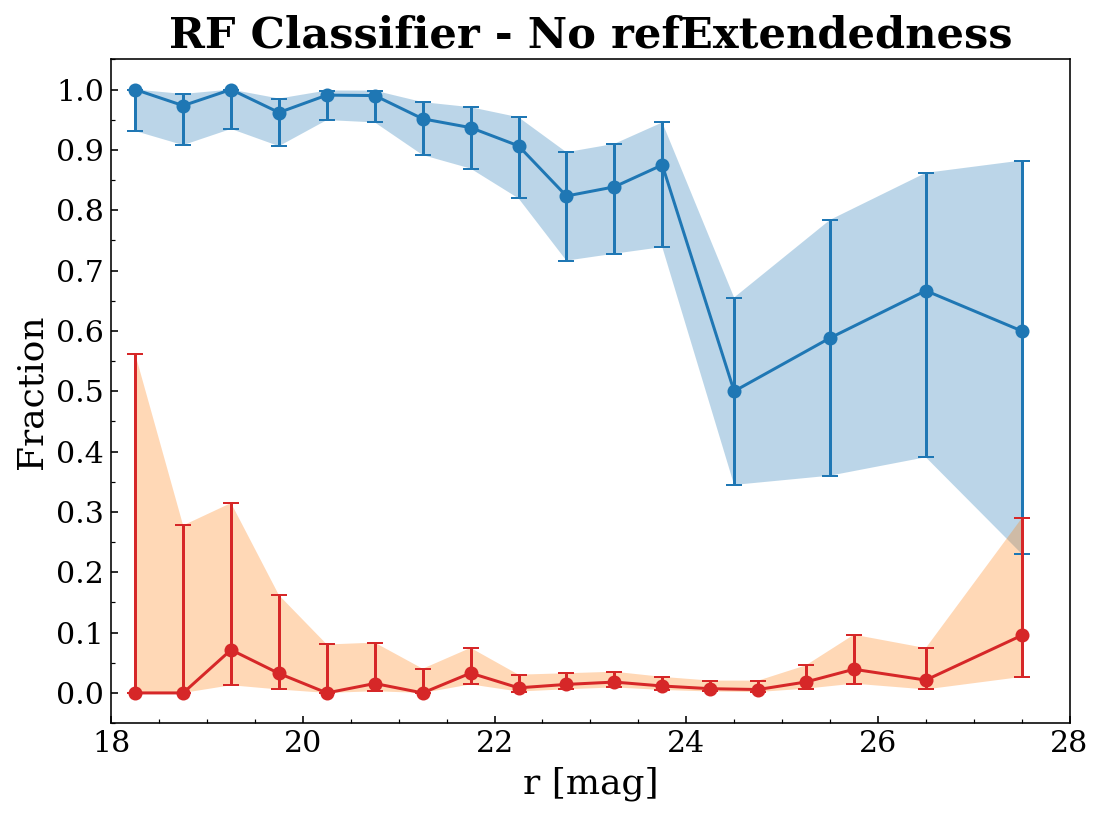}
    \includegraphics[width=.24\linewidth]{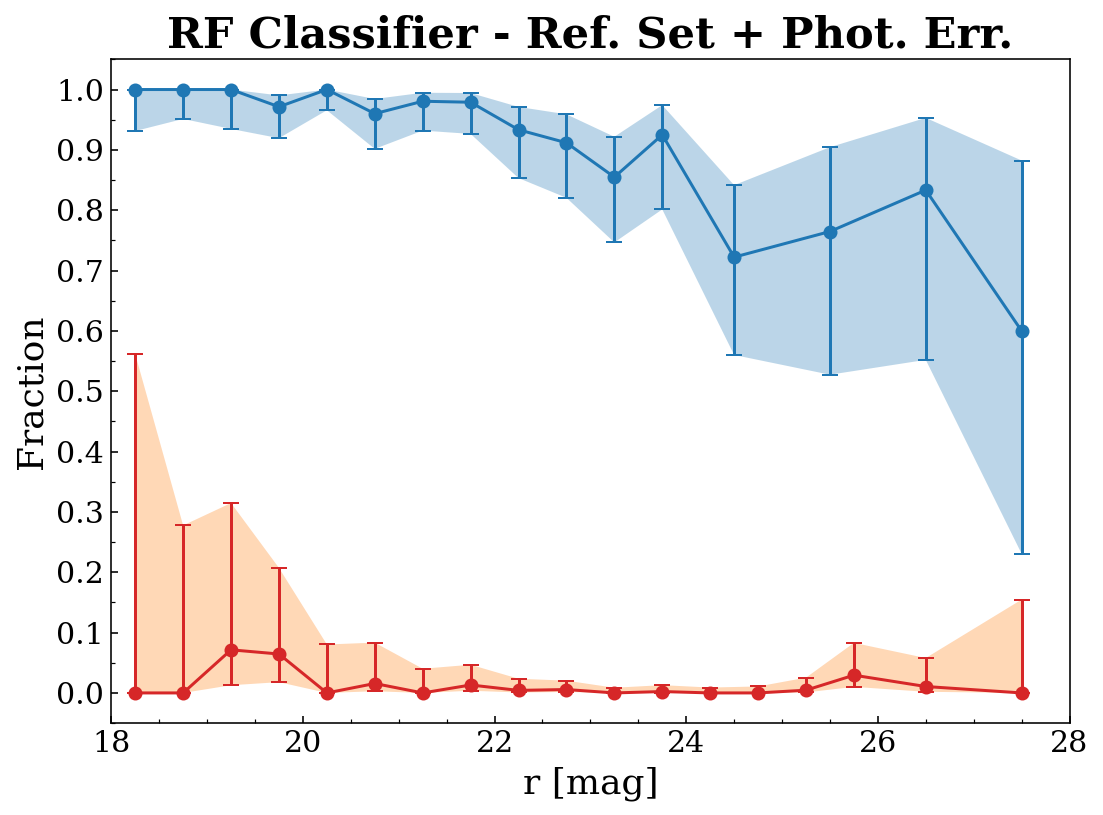}
    \caption{Performance of the {\sc refExtendedness} parameter (leftmost panel) and the random forest classifier (left-center panel), evaluated on the validation sample, as a function of the $r$-band magnitude, for the reference set. For both panels, the blue curves show the fraction of true stars correctly classified as stars (stellar completeness), while the red curves show the fraction of true galaxies misclassified as stars (galaxy contamination). The last two panels displays the performance of the random forest classifier obtained by removing the {\sc refExtendedness} parameter (right-center panel) and by adding photometric uncertainties (rightmost panel) to the reference set.}
    \label{fig:classification_fraction_result}
\end{figure*}

\subsubsection{The reference feature set: all LSST colors + {\sc refExtendedness}}

\massi{Our primary experiment adopts what we refer to as the reference feature set, composed of all possible LSST colors together with the morphological parameter {\sc refExtendedness}.
Among the 15,611 sources of our catalog, 14,360 objects, 4,002 stars (28\%) and 10,358 galaxies (72\%), have complete photometry in all six LSST filters and a valid value of the {\sc refExtendedness} parameter.
The fine-tuning analysis described in Sect.~\ref{sec:rf_tuning} revealed that six distinct combinations of hyperparameters yielded the same maximum F1 score of 93.3\%. 
All these models share the same core configuration, namely n\_estimators = 300, criterion = entropy, min\_samples\_leaf = 2, min\_samples\_split = 2, and bootstrap = False. 
The only variations occur in the choice of max\_features (either sqrt or log2), and max\_depth (None, 50 or 100).}\par 
\massi{As already anticipated in the previous sections, }
to assess the robustness and generalization capability of the classifier, we evaluated its performance on the independent validation sample, consisting of 30\% of the reference dataset (4,308 objects), which was not used during training. 
We first consider the overall accuracy, defined as:
\begin{equation}
    \rm accuracy = \frac{TP + TN}{TP + TN + FP + FN}
\end{equation}
which reaches 97.8\%. This means that, on average, the classifier correctly identifies 98 out of 100 objects. However, accuracy alone can be misleading in the presence of class imbalance, and therefore we also examine precision, recall, and F1 score separately for stars and galaxies.
Table~\ref{tab:metrics} summarizes these metrics. Both classes exhibit high precision ($\sim$98\%), indicating that when the classifier assigns an object to a given class, the classification is correct in nearly all cases. The recall is very high for galaxies (99.1\%), while slightly lower for stars (94.3\%), meaning that a small fraction of stars (about 6\%) are misclassified as galaxies. This behavior is expected given the class imbalance and the intrinsic difficulty of distinguishing faint stars from compact galaxies. \massi{While looking for UFDs this issue will slightly reduce the discovery power (since a few stars of a real UFD may be missed due to misclassification) but will not enhance false detections, that are driven by galaxies misclassified as stars, not the opposite.}
These results are also reflected in the confusion matrix, displayed in the top panel of Figure~\ref{fig:confusion_matrix}, which shows that the model correctly classifies the vast majority of galaxies (3,081 out of 3,109) and stars (1,130 out of 1,199), with most of the misclassifications occurring in the form of stars incorrectly labeled as galaxies.
Overall, the F1 score is high for both classes (98.5\% for galaxies and 95.9\% for stars), demonstrating that the classifier achieves a strong balance between completeness and purity. These results confirm that the Random Forest model is robust and effective in separating stars from galaxies with LSST photometry.\par
In Figure~\ref{fig:classification_fraction_result} we compare the performance of our Random Forest classifier (leftmost panel) with that of the {\sc refExtendedness} parameter alone (left-center panel), as a function of the $r$-band magnitude, using only objects in the validation sample. Since the leftmost panel is constructed in the same manner as Figure~\ref{fig:class_fraction_lsst} but restricted to the validation set, the overall trends remain consistent, with fluctuations due to the smaller number of objects in some magnitude bins. To reduce the statistical noise at faint magnitudes, we adopted wider magnitude bins for $r \geq 24$ mag for stars and $r \geq 26$ mag for galaxies.
As seen in the leftmost panel, the performance of {\sc refExtendedness} for stars remains high down to $r \simeq 24$ mag, after which the stellar recovery fraction decreases and fluctuates around $\sim$60\%.
Conversely, the contamination from compact galaxies misclassified as stars remains below $\sim$20\% down to $r \sim 26$ mag, but then increases sharply, exceeding $\sim$40\% by $r \sim 27.5$ mag. This behaviour mirrors what was observed in the full sample (Figure~\ref{fig:class_fraction_lsst}), confirming that {\sc refExtendedness} loses discriminating power at faint magnitudes.\par
The left-center panel demonstrates the improvement introduced by the Random Forest classifier adopting this set of features. At bright magnitudes ($r \lesssim 24$ mag), we can observe a slight improvement with respect to the {\sc refExtendedness} classifier, as expected since most bright sources are clearly resolved. However, at fainter magnitudes ($r \geq 24$ mag), the Random Forest substantially increases the stellar recovery fraction, as the latter stabilizes between 70\% and 85\%, compared to the $\sim$50\%-75\% obtained using {\sc refExtendedness} alone. 
We note, however, that the performance at these faint magnitudes is affected by small-number statistics: the validation set contains only 31 stars with $r \geq 25$ mag and 16 with $r \geq 26$ mag.\par
For galaxies, the improvement is even more pronounced. The Random Forest keeps the fraction of galaxies misclassified as stars below $\sim$5\% up to $r \sim 26.5$ mag, and only $\sim$14\% at $r \sim 27.5$ mag. This is in stark contrast with {\sc refExtendedness}, for which galaxy contamination rises rapidly beyond $r \sim 26$ mag.
These results demonstrate that the multi-band color information is the key factor enabling a significant reduction of galaxy contamination at faint magnitudes, with the Random Forest serving as an effective tool to exploit this information beyond what is achievable with morphology alone.\par

\subsubsection{Removing morphology information: LSST colors only}
We also evaluated the performance of the Random Forest classifier when the {\sc refExtendedness} parameter is excluded from the input feature set, in order to assess the capability of the model to distinguish stars and galaxies using only LSST photometry.
We repeated the optimization procedure described in Sect.~\ref{sec:rf_tuning} and found that six distinct combinations of hyperparameters yield the same maximum F1 score of 90.1\%. 
All the best–performing models share the following configuration: n\_estimators = 300, criterion = entropy, min\_samples\_leaf = 1, min\_samples\_split = 2, and bootstrap = True. 
The only differences among these top models lie in the choice of max\_features (either sqrt or log2) and in the adopted max\_depth, which can be None, 50, or 100.
The corresponding metrics are reported in the second row of Table~\ref{tab:metrics}.
The overall accuracy is slightly lower than in the case where {\sc refExtendedness} is included, reaching 96.6\%. 
The decrease primarily affects the stellar classification: while the metrics for the galaxy class degrade by less than $\sim$1\%, those for the stars worsen by about 2\%. In particular, the stellar recall drops to 91.7\%, meaning that approximately 1 out of 10 stars is misclassified as a galaxy. 
The performance as a function of magnitude, shown in the right-center panel of Figure~\ref{fig:classification_fraction_result}, confirms these trends. 
Overall, the stellar recall is lower than in the configuration including {\sc refExtendedness}, and, in particular, the random forest without the morphological parameter does not provide a clear improvement over the internal LSST stellar classification alone in terms of star recovery.
Conversely, the galaxy contamination fraction obtained with the random forest remains comparable between this experiment and the previous one. These results show that LSST multi-band color information alone is already sufficient to   keep the contamination from background galaxies at very low levels across the full magnitude range probed, including the faint end where morphological information becomes progressively ineffective. 
At the same time, the absence of the morphological parameter leads to a reduced recovery of stars, which becomes more evident at intermediate and faint magnitudes ($r \gtrsim 24$ mag). In this regime, {\sc refExtendedness} still provides a measurable benefit in terms of stellar recall, even though its overall discriminating power is diminished.
We note, however, that the stellar validation sample at these magnitudes is affected by low-number statistics. An expanded training and validation set with securely identified faint stars may therefore improve the stellar recall even in configurations that rely exclusively on multi-band photometric information.
Interestingly, part of this loss in stellar recall can be mitigated when photometric uncertainties are explicitly included among the input features, even without the use of {\sc refExtendedness}. 
The detailed results of this experiment are discussed in~\ref{app:photoerr_only}.

\subsubsection{Adding photometric uncertainties}

In this third experiment, we augmented the reference feature set (all LSST colors + \Extend~) by adding the photometric uncertainties in each filter. The motivation behind this test is to evaluate whether incorporating measurement errors can improve classification performance at faint magnitudes, where photometric noise becomes significant and sources of different classes tend to scatter into overlapping regions of color–color space (see also Fig.~\ref{fig:color_color}).
A new hyperparameter optimization following the procedure described in Sect.~\ref{sec:rf_tuning} identified two distinct configurations achieving the same maximum F1 score of 93.5\%, the highest obtained among the three experiments discussed so far. Both optimal models share the following parameters: n\_estimators = 1000, criterion = gini, min\_samples\_leaf = 1, min\_samples\_split = 2, max\_depth = 10, and bootstrap = True, differing only in the choice of max\_features (either sqrt or log2).\par
The performance metrics reported in the third row of Table~\ref{tab:metrics} confirm that this configuration yields the best overall results. The accuracy reaches 98.3\%, outperforming both the reference model and all other feature combinations. While the improvement for galaxies is modest (a few tenths of a percent across precision, recall, and F1 score), the gain for stars is more substantial: stellar precision increases to almost 99\%, recall to $\simeq$ 95\%, and F1 score to $\simeq$97\%.
These improvements are also evident in the confusion matrix (bottom panel of Fig.~\ref{fig:confusion_matrix}), where the number of misclassified galaxies is reduced by almost a factor of two, and the number of misclassified stars decreases by $\simeq$15\% compared to the reference set. 
The magnitude-dependent performance (rightmost panel of Fig.~\ref{fig:classification_fraction_result}) shows that galaxy contamination remains negligible across the entire magnitude range probed, while the stellar recall is comparable (within uncertainties) to that of the reference configuration.
Overall, these results indicate that incorporating photometric uncertainties provides valuable additional information to the classifier, enabling it to better account for the scatter of faint sources away from their intrinsic loci in color–color space. This effect is particularly important for stars, whose locus is defined by narrow sequence \citep[e.g.][and see also Fig.~\ref{fig:color_color}]{Fadely-2012}; poorly measured stars are more likely to deviate from this tight sequence, and including their uncertainties helps the classifier correctly recover them. Conversely, galaxies occupy a broader and more diffuse region of color–color space, so the benefit for this class is more limited, but still appreciable.

\subsubsection{Key findings from the feature set tests}

To summarize the results of these three experiments, our analysis shows that, for the present sample of bona-fide stars and galaxies, LSST multi-band colors are sufficient to maintain negligible galaxy contamination across the entire magnitude range (e.g. $r \lesssim$ 27.5 mag). While the inclusion of {\sc refExtendedness} further enhances stellar recall, particularly at the faintest magnitudes, it offers no additional gain in reducing galaxy misclassification, as evidenced by the consistently low contamination levels even in the absence of morphological data  (right-center panel of Fig.~\ref{fig:classification_fraction_result}).
Finally, incorporating photometric uncertainties as input features enhances the overall performance of the random forest classifier, especially for stars, whose large color errors can otherwise scatter them away from the narrow stellar locus in color–color diagrams.\par
We performed an additional test using the XGBoost algorithm \citep{XGBoost-2016, XGBoost-2026}, which is widely adopted in classification tasks and is known for its computational efficiency. 
The performance metrics obtained with XGBoost are widely consistent with those of Random Forest, with marginally lower metrics and a slightly higher contamination at the faint end. Given our primary scientific objective of minimizing galaxy contamination in stellar catalogs, we retain the Random Forest classifier as our reference model. A detailed comparison is provided in Appendix~\ref{app:xgboost}.

\subsection{Feature importance}

\begin{figure}
    \centering
    \includegraphics[width=.9\linewidth]{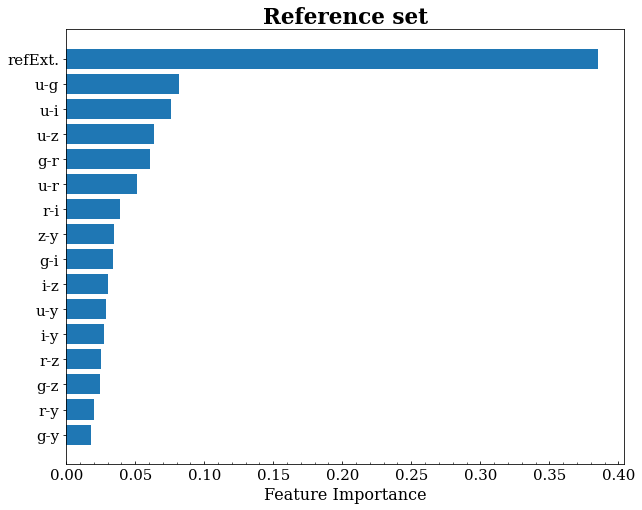}\\
    \includegraphics[width=.9\linewidth]{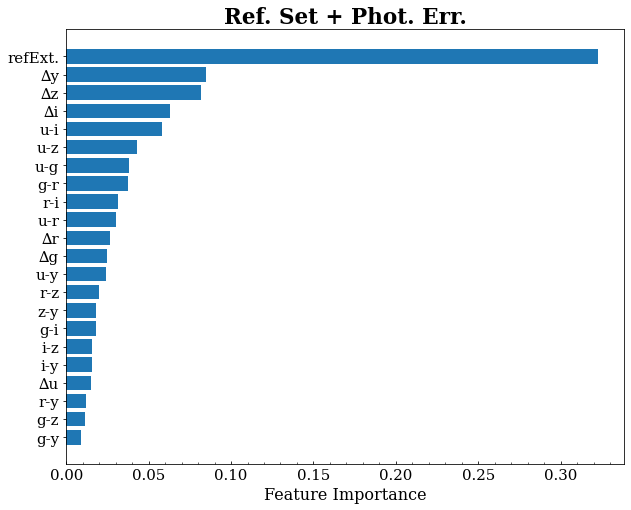}
    \caption{Relative importance of each feature \massi{for the reference set (top panel) and with the inclusion of photometric uncertainties (bottom panel).}}
    \label{fig:feature_importance}
\end{figure}

A key advantage of Random Forest classifiers is their ability to quantify the relative importance of the input features used during the classification process.
We computed the feature importance values using the {\it feature\_importances\_} attribute provided by {\it scikit-learn}, which measures the average reduction in node impurity contributed by each feature across all decision trees in the ensemble.
The top panel of Figure~\ref{fig:feature_importance} displays the resulting ranking of features, from the most to the least relevant, by adopting the reference set of features.
The {\sc refExtendedness} parameter clearly dominates the feature space, contributing alone to nearly 40\% of the total importance.
This strong discriminating power is already evident at bright magnitudes (Figure~\ref{fig:class_fraction_lsst}), where morphological information efficiently separates stars from galaxies.
The dominance of the morphological parameter in our feature-importance analysis is consistent with previous studies. For example, \citet{Jeakel-2026} showed that the most informative predictors in their XGBoost model are the concentration index and the normalized peak surface brightness (see their figure 17), while \citet{Khramtsov-2019} found that the KiDS {\sc CLASS\_STAR} parameter overwhelmingly dominates the information content (their figure 4). It is worth noting, however, that both works operate at substantially shallower depth than the LSST regime explored here.
Indeed, in our experiments, as both Figure~\ref{fig:class_fraction_lsst} and the leftmost panel of Figure~\ref{fig:classification_fraction_result} demonstrate, the effectiveness of {\sc refExtendedness} alone decreases at intermediate and faint magnitudes.
In this regime, where morphological information becomes progressively less reliable, multi-band color information plays an increasingly dominant role. Indeed, at the faintest magnitudes probed in this work, the classifier achieves comparable, or even slightly improved, performance, in terms of galaxy classification, when {\sc refExtendedness} is excluded from the feature set (see the right-center panel of Figure~\ref{fig:classification_fraction_result}). This indicates that LSST multi-band photometry alone already carries most of the discriminative power required to control galaxy contamination in the faint regime.
It is important to note that the feature-importance analysis presented here is global, i.e. it does not explicitly account for magnitude-dependent effects. As a result, parameters that are highly effective at bright magnitudes, such as {\sc refExtendedness}, naturally dominate the ranking, even though their discriminating power decreases toward the faint end.
Consistently with this interpretation, we find that the reduced stellar recall observed at faint magnitudes when excluding {\sc refExtendedness} can be largely recovered by incorporating photometric uncertainties into the feature set (see~\ref{app:photoerr_only}), indicating that color information, when properly weighted by its uncertainties, is sufficient to drive the classification in the low signal-to-noise regime.\par
Among the color indices, the most relevant feature is $(u-g)$, which contributes almost 10\% of the total importance.
Other significant features include $(u-i)$, $(u-z)$, $(g-r)$, and $(u-r)$, each contributing more than 5\%.
The importance then gradually decreases for the remaining color combinations, all below the 5\%.
Overall, the $u$-band emerges as a key player in the classification, being involved in nearly all the top-ranked color indices.
It is worth noting that this trend is broadly consistent with previous studies. For instance, \citet{Khramtsov-2019} found that the most discriminating color for their CatBoost classifier was the near-infrared index $(H-K_s)$, followed by $(u-g)$ as the second most informative feature. This result highlights two complementary points: 
(i) extending the wavelength coverage deeper into the near-infrared (beyond LSST’s $y$ band) can significantly enhance star–galaxy separation, suggesting that infrared imaging from missions such as Euclid (though lacking $K_s$, but including $J$ and $H$) may provide valuable additional leverage;
(ii) in the absence of such deep infrared information, the $(u-g)$ color remains the strongest optical discriminant, fully consistent with what we observe in our feature set.
Finally, apart from the presence of the $z$-filter in the third-ranked color combination, the LSST infrared bands contribute relatively little to the classification task.
This indicates that the combination of {\sc refExtendedness} and $u$-band colors already captures most of the discriminative information needed, reducing the relative impact of redder bands in this particular feature set.\par
The bottom panel of Figure~\ref{fig:feature_importance} displays the feature importance ranking obtained when photometric uncertainties are added to the reference feature set.
Interestingly, immediately after \Extend, which remains the single most informative feature, photometric uncertainties, particularly those in the $y$, $z$, and $i$ bands, rank from second to fourth in importance.
We interpret this result in light of the relative depths of the LSST filters in DP1: the $z$ and $y$ bands are shallower by approximately 1–3 magnitudes compared to $g$ and $r$ \citep[i.e., $g = 26.18$ mag, $r = 25.96$ mag, $z = 25.07$ mag, $y = 23.10$ mag;][]{VRO-DP1}. As a consequence, for color combinations involving these redder bands, photometric uncertainties carry significant information that helps the classifier correctly assign sources to their respective classes, particularly for stars, which occupy intrinsically narrow loci in color–color space (see Sect.~\ref{sec:color-color}). Consistent with this interpretation, the photometric uncertainties in the deeper $g$ and $r$ bands do not appear among the highest-ranked features.
Interestingly, the uncertainty in the $u$ band lies among the least informative features, despite the $u$ band being shallower than $z$ (i.e., $u = 24.55$ mag). This likely reflects the fact that color–color diagrams involving the $u$ band already provide strong intrinsic separation between stars and galaxies (see Sect.~\ref{sec:color-color}), such that additional information from photometric uncertainties does not substantially improve the classification in this case.
Indeed, even in this extended feature set, colors involving the $u$ filter (e.g., $u-i$, $u-z$, and $u-g$) remain among the most discriminative photometric features, further highlighting the critical role of the $u$ band in star–galaxy separation tasks at faint magnitudes.\par

\section{Stellar and galaxy loci in LSST color–color diagrams}
\label{sec:color-color}

\begin{figure*}
    \includegraphics[width=.24\linewidth]{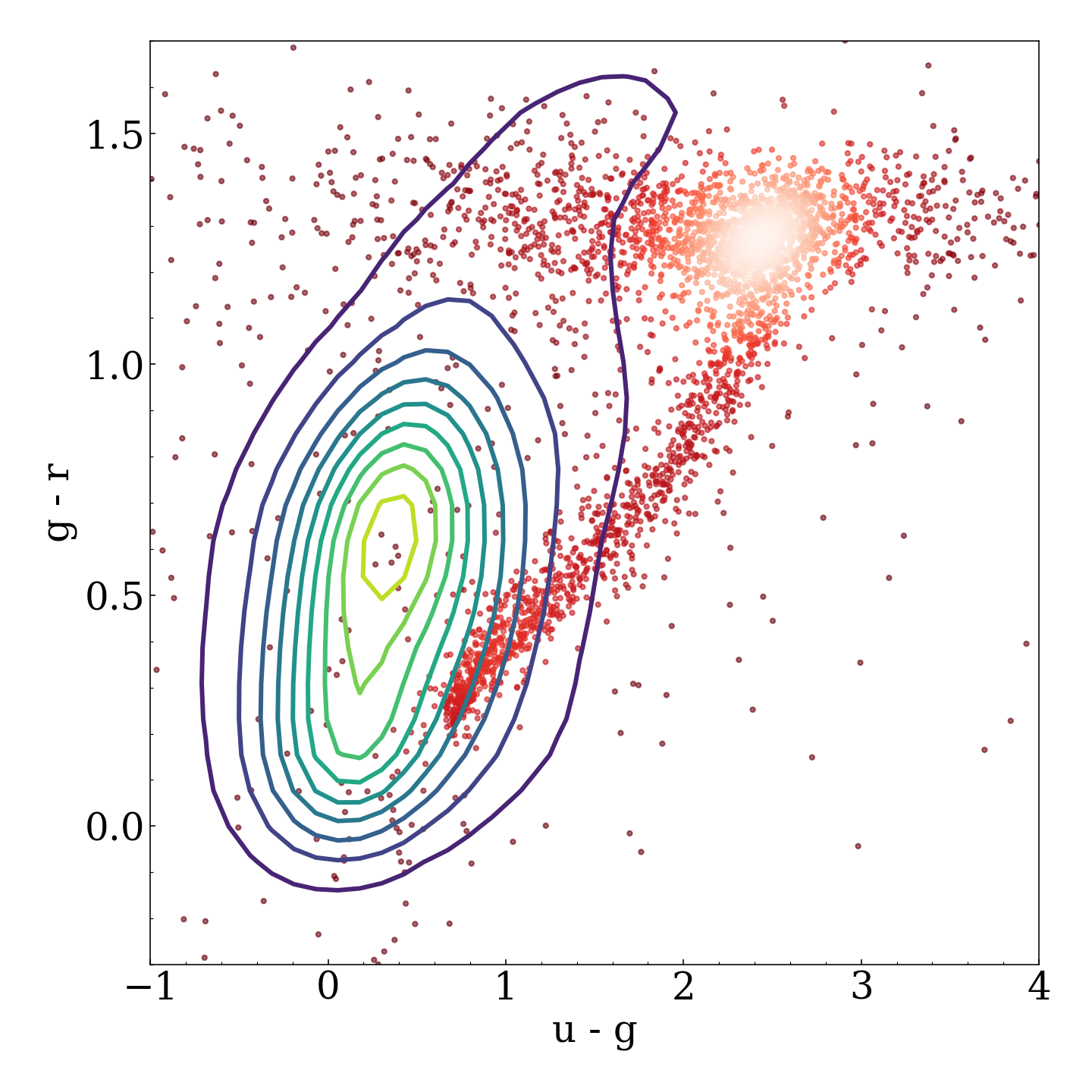}
    \includegraphics[width=.24\linewidth]{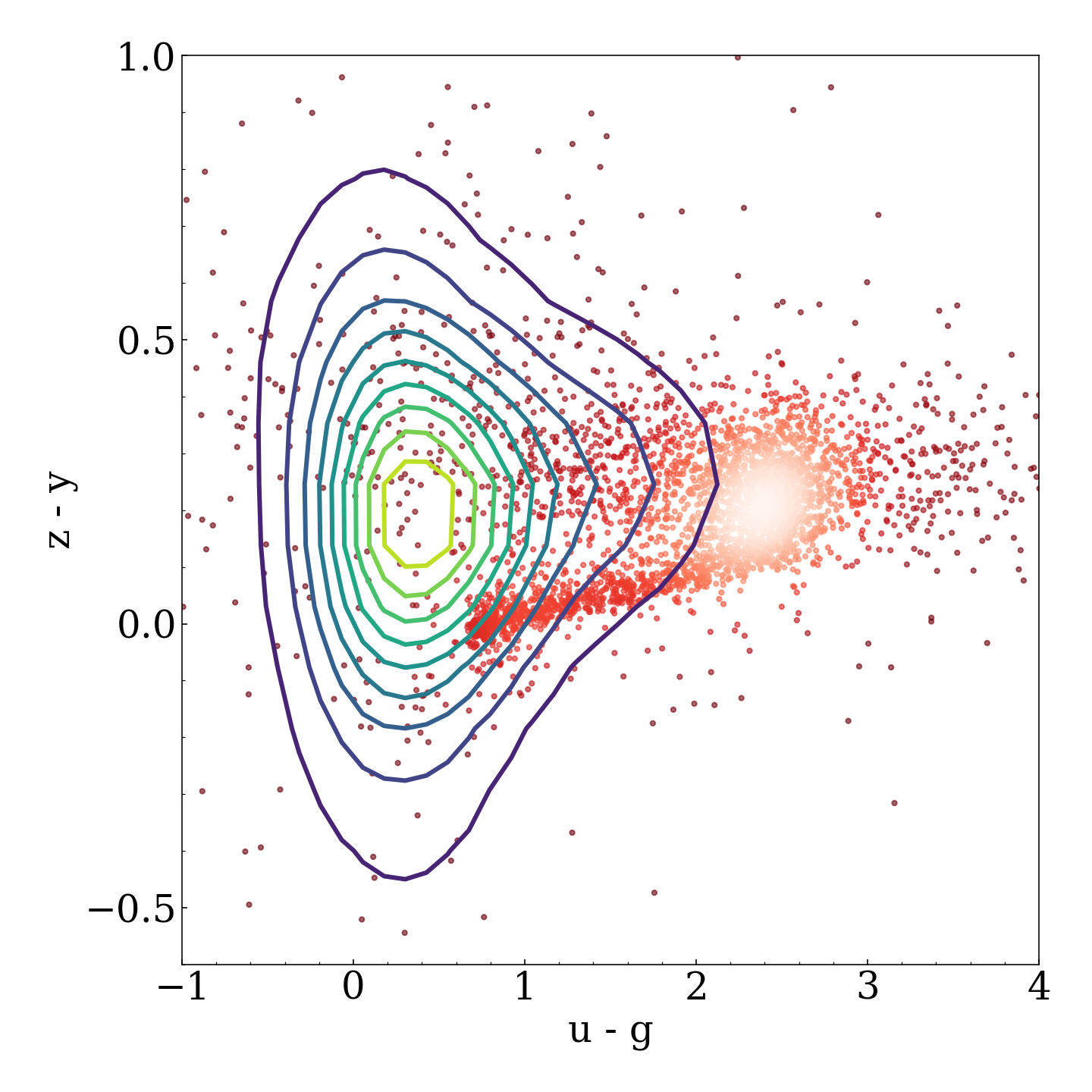}
    \includegraphics[width=.24\linewidth]{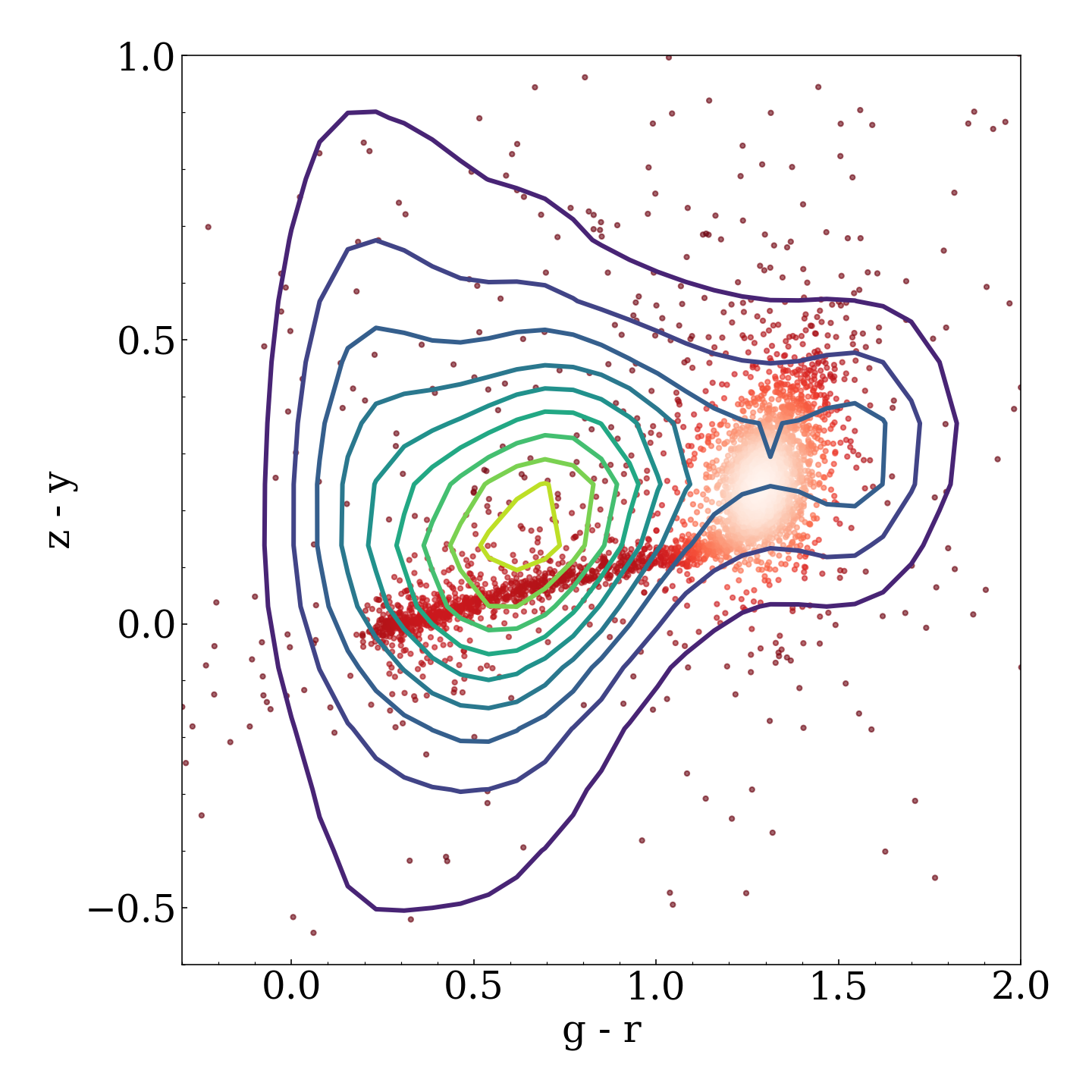}
    \includegraphics[width=.24\linewidth]{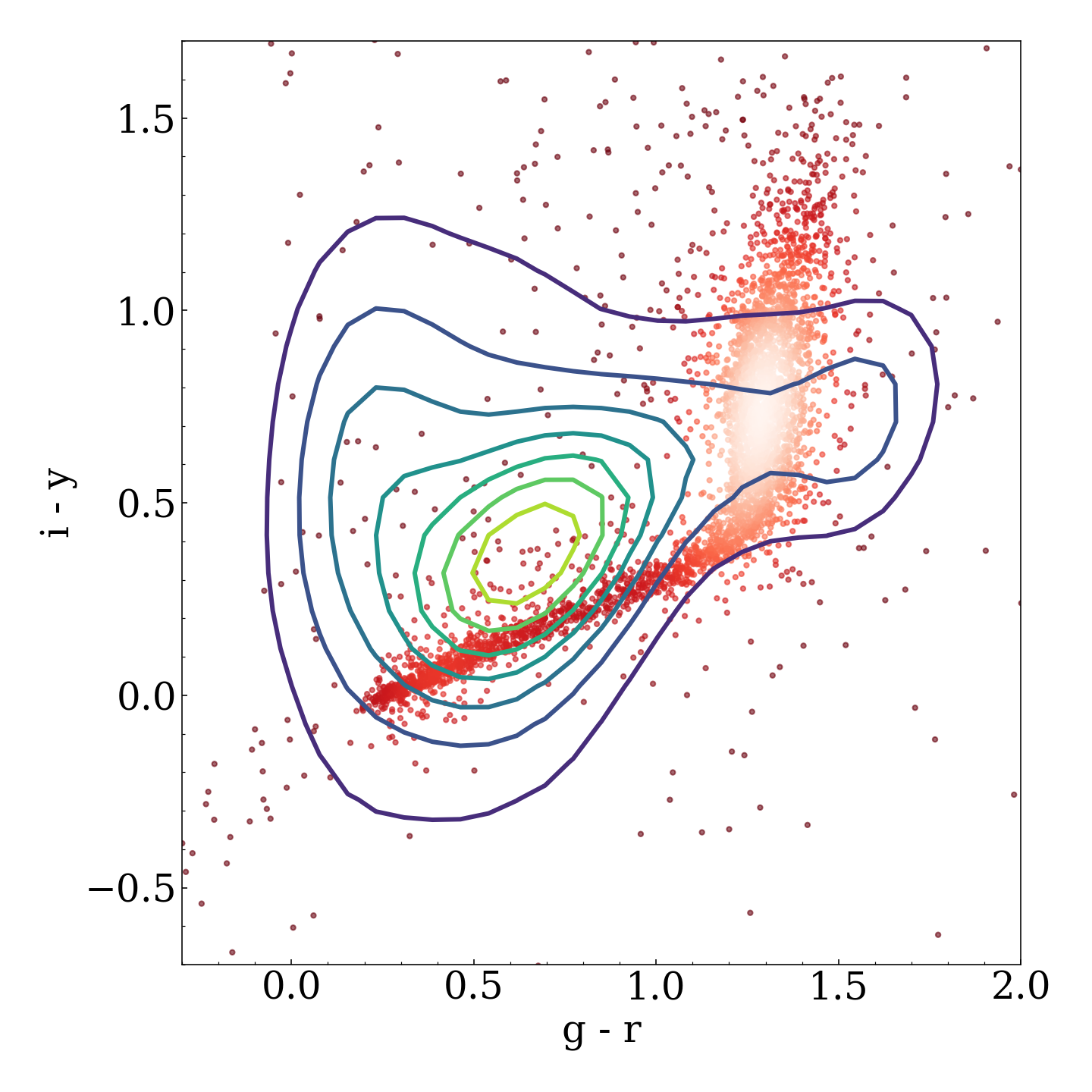}
    \caption{Color–color diagram showing the distribution of stars and galaxies in our sample for different combination of color pairs. Stellar sources (in red) are displayed as individual points color-coded by their local density.
Galaxies are represented by isodensity contours (in viridis),  highlighting the underlying structure of the galaxy locus in color–color space.}
    \label{fig:color_color}
\end{figure*}

Figure~\ref{fig:color_color} presents a set of color–color diagrams where the separation between the two populations becomes more evident. The leftmost panel shows $(u-g)$ versus $(g-r)$, the two most discriminant color features of the classifier. In this diagram, nearly all stars (red points) occupy a well-defined region distinct from galaxies. Specifically, the galaxy distribution, represented by isodensity contours, is largely confined to $-0.5 \leq (u-g) \leq 1.0$ and $0 \leq (g-r) \leq 1.0$, whereas stars follow a narrow curved sequence extending from $(u-g, g-r) \simeq (0.7, 0.2)$ to $(2.4, 1.3)$, where thick- and thin-disk MW stars converge into a compact clump \citep[see also][for a similar plot made with $\sim$2 million stars from SDSS Data Release 1]{Smolcic-2004}.
The combination of $(u-g)$ with infrared colors also provides a clear separation. For instance, the left-center panel shows $(u-g)$ versus $(z-y)$, where stars trace a narrow diagonal sequence from $(u-g, z-y) \simeq (0.7, 0.0)$ to $(2.4, 0.2)$, again terminating in a clump associated with thick- and thin-disk MW stars. Most of the galaxies, by contrast, are displaced to the left of this sequence. Similar diagrams can be obtained using $(u-r)$, or $u-i$, instead of $(u-g)$; in these cases, all sources (including galaxies) are shifted toward redder colors, with the MW stellar clump located at $u-r \simeq 3.7$ mag, or $u-i \simeq 4.7$ mag, respectively.\par
Even without the $u$ filter, certain color combinations could be more effective in separating stars from galaxies, particularly those involving optical–infrared pairs. The last two panels show $(g-r)$ versus $(z-y)$ (right-center panel) and $(i-y)$ (rightmost panel). 
In both cases, the stellar locus forms a narrow, well-defined diagonal sequence, while the galaxy population occupies a broader region of the diagram. Nonetheless, despite its compactness, this stellar sequence partially overlaps the region where the galaxy density increases significantly, around $(g-r, z-y) \simeq (0.6, 0.1)$ and $(g-r, i-y) \simeq (0.6, 0.2)$. 
This overlap is particularly problematic because the same color range ($g-r \lesssim 1.0$ mag) is where old, metal-poor stars characteristic of UFDs are found, hampering detection algorithms that rely on isolating these populations \citep[e.g.][]{Walsh-2009}.
Moreover, the stellar clump associated with thick- and thin-disk MW stars itself intersects zones of the diagram that are populated, albeit at relatively low density, by compact galaxies, leading inevitably to some degree of contamination.
These color–color diagrams therefore illustrate that in the absence of deeper near-infrared information \citep[e.g., $J - K_s$, see][their figure 1]{Tortora-2018}, which would more effectively separate stars and galaxies, optical–infrared LSST colors alone cannot fully disentangle the overlapping loci of stars and galaxies. Consequently, the inclusion of the $u$-band becomes particularly advantageous for enhancing star–galaxy separability.\par
{As an additional robustness check, we verified that the machine-learning classification preserves the intrinsic stellar locus in color–color space by directly comparing the predicted stellar distribution with that defined by our sample of confirmed stars. The two loci show a good agreement. The details of this comparison are presented in Appendix~\ref{app:color_color_true_pred}.}

\section{Summary}
\label{sec:discussion}

In this pilot study, we investigated the performance of supervised machine learning techniques, specifically a Random Forest classifier, for separating stars from galaxies in deep, LSST data. Using observations of the ECDFS from LSST DP1, we constructed a heterogeneous sample of bona fide stars and galaxies that approaches, though does not fully reach, the expected 10-year LSST depth, and whose median seeing is worse than the survey’s design goals. Consequently, the results presented here should be regarded as a conservative lower limit on the classification performance achievable with the final LSST dataset.\par
Our analysis demonstrates that incorporating LSST multi-band photometry substantially improves star–galaxy separation compared to using morphological information alone, especially at faint magnitudes, the regime in which LSST’s discovery potential will be greatest. The study also highlights that while {\sc refExtendedness} is the most informative parameter, its discriminating power rapidly declines toward the faint end. In contrast, color information, particularly $u$-band colors, provides a robust classification even in the faint regime, when morphological separation alone becomes ineffective.
Moreover, providing full six-band LSST photometry is more effective for ensuring reliable star–galaxy separation than substituting a filter with the \Extend~parameter, particularly when high stellar purity is a priority (see App.~\ref{app:rf_other_config}).
Finally, explicitly including photometric uncertainties as input features provides additional discriminative power, especially at faint magnitudes where increased photometric scatter causes sources to migrate across color–color loci. This leads to a measurable improvement in stellar classification, and a slight decreasing in galaxy contamination as well.
This confirms that information from multi-band photometry is essential for reliable star–galaxy separation in LSST-like data.\par
{A comparison with the XGBoost algorithm shows fully consistent results, with marginally lower performance. Specifically, it shows a slight increased contamination at the faint end, with more galaxies misclassified as stars at the last magnitude bins.}
Overall, our results demonstrate that supervised machine-learning methods can keep galaxy contamination at negligible levels down to very faint magnitudes, ensuring that searches for UFDs will not be significantly hindered by clustered misclassified background galaxies.
A limitation instead might lie in stellar completeness, particularly due to the scarcity of well-characterized faint stars in current training sets.
These findings provide a solid foundation for optimizing classification strategies in LSST and other multi-band surveys. Further performance gains will likely require access to larger, reliable training samples that adequately cover the faint end of the magnitude distribution, an observationally challenging but worthwhile goal.
To support future work in this direction, we will publicly release the curated star–galaxy sample assembled for this study, enabling the community to extend, and refine classification models as LSST data quality and depth improve.

\section*{Data availability}

The catalog of bona-fide stars and galaxies built in this work is available at

\begin{acknowledgements}


In this work, we made use of the following softwares: 
TOPCAT \citep{topcat}, Scikit-learn \citep{scikit-learn}, NumPy \citep{numpy}, pandas \citep{pandas-2010, pandas-2020}, matplotlib \citep{matplotlib}, and XGBoost \citep{XGBoost-2016, XGBoost-2026}.\par
We thank the anonymous referee for the constructive comments, which have significantly improved the quality of the manuscript.\\
This work has made use of data from the European Space Agency (ESA) mission
{\it Gaia} (\url{https://www.cosmos.esa.int/gaia}), processed by the {\it Gaia}
Data Processing and Analysis Consortium (DPAC,
\url{https://www.cosmos.esa.int/web/gaia/dpac/consortium}). Funding for the DPAC
has been provided by national institutions, in particular the institutions
participating in the {\it Gaia} Multilateral Agreement.\\
This work is based on observations taken by the 3D-HST Treasury Program (GO 12177 and 12328) with the NASA/ESA HST, which is operated by the Association of Universities for Research in Astronomy, Inc., under NASA contract NAS5-26555.\\
This work is based on observations taken by the CANDELS Multi-Cycle Treasury Program with the NASA/ESA HST, which is operated by the Association of Universities for Research in Astronomy, Inc., under NASA contract NAS5-26555.\\
This research used data obtained with the Dark Energy Spectroscopic Instrument (DESI). DESI construction and operations is managed by the Lawrence Berkeley National Laboratory. This research is supported by the U.S. Department of Energy, Office of Science, Office of High-Energy Physics, under Contract No. DE–AC02–05CH11231, and by the National Energy Research Scientific Computing Center, a DOE Office of Science User Facility under the same contract. Additional support for DESI is provided by the U.S. National Science Foundation, Division of Astronomical Sciences under Contract No. AST-0950945 to the NSF’s National Optical-Infrared Astronomy Research Laboratory; the Science and Technology Facilities Council of the United Kingdom; the Gordon and Betty Moore Foundation; the Heising-Simons Foundation; the French Alternative Energies and Atomic Energy Commission (CEA); the National Council of Science and Technology of Mexico (CONACYT); the Ministry of Science and Innovation of Spain, and by the DESI Member Institutions. The DESI collaboration is honored to be permitted to conduct astronomical research on Iolkam Du’ag (Kitt Peak), a mountain with particular significance to the Tohono O’odham Nation.\\
We thank M. D'addona for his valuable insights on the random forest classifier.\\
M.G. acknowledges "Partecipazione LSST - Large Synoptic Survey Telescope (ref. A. Fontana)" (Ob. Fu.: 1.05.03.06).

\end{acknowledgements}

%
\bibliographystyle{aa} 
\bibliography{RandomForest} 

@ARTICLE{Balestra-2010,
       author = {{Balestra}, I. and {Mainieri}, V. and {Popesso}, P. and {Dickinson}, M. and {Nonino}, M. and {Rosati}, P. and {Teimoorinia}, H. and {Vanzella}, E. and {Cristiani}, S. and {Cesarsky}, C. and {Fosbury}, R.~A.~E. and {Kuntschner}, H. and {Rettura}, A.},
        title = "{The Great Observatories Origins Deep Survey. VLT/VIMOS spectroscopy in the GOODS-south field: Part II}",
      journal = {\aap},
         year = 2010,
        month = mar,
       volume = {512},
          eid = {A12},
        pages = {A12},
          doi = {10.1051/0004-6361/200913626},
archivePrefix = {arXiv},
       eprint = {1001.1115},
 primaryClass = {astro-ph.CO},
       adsurl = {https://ui.adsabs.harvard.edu/abs/2010A&A...512A..12B}
}

@ARTICLE{Vanzella-2008,
       author = {{Vanzella}, E. and {Cristiani}, S. and {Dickinson}, M. and {Giavalisco}, M. and {Kuntschner}, H. and {Haase}, J. and {Nonino}, M. and {Rosati}, P. and {Cesarsky}, C. and {Ferguson}, H.~C. and {Fosbury}, R.~A.~E. and {Grazian}, A. and {Moustakas}, L.~A. and {Rettura}, A. and {Popesso}, P. and {Renzini}, A. and {Stern}, D. and {GOODS Team}},
        title = "{The great observatories origins deep survey. VLT/FORS2 spectroscopy in the GOODS-South field: Part III}",
      journal = {\aap},
         year = 2008,
        month = jan,
       volume = {478},
       number = {1},
        pages = {83-92},
          doi = {10.1051/0004-6361:20078332},
archivePrefix = {arXiv},
       eprint = {0711.0850},
 primaryClass = {astro-ph},
       adsurl = {https://ui.adsabs.harvard.edu/abs/2008A&A...478...83V}
}

@ARTICLE{Popesso-2009,
       author = {{Popesso}, P. and {Dickinson}, M. and {Nonino}, M. and {Vanzella}, E. and {Daddi}, E. and {Fosbury}, R.~A.~E. and {Kuntschner}, H. and {Mainieri}, V. and {Cristiani}, S. and {Cesarsky}, C. and {Giavalisco}, M. and {Renzini}, A. and {GOODS Team}},
        title = "{The great observatories origins deep survey. VLT/VIMOS spectroscopy in the GOODS-south field}",
      journal = {\aap},
         year = 2009,
        month = feb,
       volume = {494},
       number = {2},
        pages = {443-460},
          doi = {10.1051/0004-6361:200809617},
archivePrefix = {arXiv},
       eprint = {0802.2930},
 primaryClass = {astro-ph},
       adsurl = {https://ui.adsabs.harvard.edu/abs/2009A&A...494..443P}
}

@ARTICLE{Momcheva-2016,
       author = {{Momcheva}, Ivelina G. and {Brammer}, Gabriel B. and {van Dokkum}, Pieter G. and {Skelton}, Rosalind E. and {Whitaker}, Katherine E. and {Nelson}, Erica J. and {Fumagalli}, Mattia and {Maseda}, Michael V. and {Leja}, Joel and {Franx}, Marijn and {Rix}, Hans-Walter and {Bezanson}, Rachel and {Da Cunha}, Elisabete and {Dickey}, Claire and {F{\"o}rster Schreiber}, Natascha M. and {Illingworth}, Garth and {Kriek}, Mariska and {Labb{\'e}}, Ivo and {Ulf Lange}, Johannes and {Lundgren}, Britt F. and {Magee}, Daniel and {Marchesini}, Danilo and {Oesch}, Pascal and {Pacifici}, Camilla and {Patel}, Shannon G. and {Price}, Sedona and {Tal}, Tomer and {Wake}, David A. and {van der Wel}, Arjen and {Wuyts}, Stijn},
        title = "{The 3D-HST Survey: Hubble Space Telescope WFC3/G141 Grism Spectra, Redshifts, and Emission Line Measurements for \raisebox{-0.5ex}\textasciitilde 100,000 Galaxies}",
      journal = {\apjs},
         year = 2016,
        month = aug,
       volume = {225},
       number = {2},
          eid = {27},
        pages = {27},
          doi = {10.3847/0067-0049/225/2/27},
archivePrefix = {arXiv},
       eprint = {1510.02106},
 primaryClass = {astro-ph.GA},
       adsurl = {https://ui.adsabs.harvard.edu/abs/2016ApJS..225...27M}
}

@ARTICLE{Mutlu-Pakdil-2021,
       author = {{Mutlu-Pakdil}, Bur{\c{c}}in and {Sand}, David J. and {Crnojevi{\'c}}, Denija and {Drlica-Wagner}, Alex and {Caldwell}, Nelson and {Guhathakurta}, Puragra and {Seth}, Anil C. and {Simon}, Joshua D. and {Strader}, Jay and {Toloba}, Elisa},
        title = "{Resolved Dwarf Galaxy Searches within  5 Mpc with the Vera Rubin Observatory and Subaru Hyper Suprime-Cam}",
      journal = {\apj},
         year = 2021,
        month = sep,
       volume = {918},
       number = {2},
          eid = {88},
        pages = {88},
          doi = {10.3847/1538-4357/ac0db8},
archivePrefix = {arXiv},
       eprint = {2105.01658},
 primaryClass = {astro-ph.GA},
       adsurl = {https://ui.adsabs.harvard.edu/abs/2021ApJ...918...88M}
}

@ARTICLE{Ivezic-2019,
       author = {{Ivezi{\'c}}, {\v{Z}}eljko and {Kahn}, Steven M. and {Tyson}, J. Anthony and {Abel}, Bob and {Acosta}, Emily and {Allsman}, Robyn and {Alonso}, David and {AlSayyad}, Yusra and {Anderson}, Scott F. and {Andrew}, John and {Angel}, James Roger P. and {Angeli}, George Z. and {Ansari}, Reza and {Antilogus}, Pierre and {Araujo}, Constanza and {Armstrong}, Robert and {Arndt}, Kirk T. and {Astier}, Pierre and {Aubourg}, {\'E}ric and {Auza}, Nicole and {Axelrod}, Tim S. and {Bard}, Deborah J. and {Barr}, Jeff D. and {Barrau}, Aurelian and {Bartlett}, James G. and {Bauer}, Amanda E. and {Bauman}, Brian J. and {Baumont}, Sylvain and {Bechtol}, Ellen and {Bechtol}, Keith and {Becker}, Andrew C. and {Becla}, Jacek and {Beldica}, Cristina and {Bellavia}, Steve and {Bianco}, Federica B. and {Biswas}, Rahul and {Blanc}, Guillaume and {Blazek}, Jonathan and {Blandford}, Roger D. and {Bloom}, Josh S. and {Bogart}, Joanne and {Bond}, Tim W. and {Booth}, Michael T. and {Borgland}, Anders W. and {Borne}, Kirk and {Bosch}, James F. and {Boutigny}, Dominique and {Brackett}, Craig A. and {Bradshaw}, Andrew and {Brandt}, William Nielsen and {Brown}, Michael E. and {Bullock}, James S. and {Burchat}, Patricia and {Burke}, David L. and {Cagnoli}, Gianpietro and {Calabrese}, Daniel and {Callahan}, Shawn and {Callen}, Alice L. and {Carlin}, Jeffrey L. and {Carlson}, Erin L. and {Chandrasekharan}, Srinivasan and {Charles-Emerson}, Glenaver and {Chesley}, Steve and {Cheu}, Elliott C. and {Chiang}, Hsin-Fang and {Chiang}, James and {Chirino}, Carol and {Chow}, Derek and {Ciardi}, David R. and {Claver}, Charles F. and {Cohen-Tanugi}, Johann and {Cockrum}, Joseph J. and {Coles}, Rebecca and {Connolly}, Andrew J. and {Cook}, Kem H. and {Cooray}, Asantha and {Covey}, Kevin R. and {Cribbs}, Chris and {Cui}, Wei and {Cutri}, Roc and {Daly}, Philip N. and {Daniel}, Scott F. and {Daruich}, Felipe and {Daubard}, Guillaume and {Daues}, Greg and {Dawson}, William and {Delgado}, Francisco and {Dellapenna}, Alfred and {de Peyster}, Robert and {de Val-Borro}, Miguel and {Digel}, Seth W. and {Doherty}, Peter and {Dubois}, Richard and {Dubois-Felsmann}, Gregory P. and {Durech}, Josef and {Economou}, Frossie and {Eifler}, Tim and {Eracleous}, Michael and {Emmons}, Benjamin L. and {Fausti Neto}, Angelo and {Ferguson}, Henry and {Figueroa}, Enrique and {Fisher-Levine}, Merlin and {Focke}, Warren and {Foss}, Michael D. and {Frank}, James and {Freemon}, Michael D. and {Gangler}, Emmanuel and {Gawiser}, Eric and {Geary}, John C. and {Gee}, Perry and {Geha}, Marla and {Gessner}, Charles J.~B. and {Gibson}, Robert R. and {Gilmore}, D. Kirk and {Glanzman}, Thomas and {Glick}, William and {Goldina}, Tatiana and {Goldstein}, Daniel A. and {Goodenow}, Iain and {Graham}, Melissa L. and {Gressler}, William J. and {Gris}, Philippe and {Guy}, Leanne P. and {Guyonnet}, Augustin and {Haller}, Gunther and {Harris}, Ron and {Hascall}, Patrick A. and {Haupt}, Justine and {Hernandez}, Fabio and {Herrmann}, Sven and {Hileman}, Edward and {Hoblitt}, Joshua and {Hodgson}, John A. and {Hogan}, Craig and {Howard}, James D. and {Huang}, Dajun and {Huffer}, Michael E. and {Ingraham}, Patrick and {Innes}, Walter R. and {Jacoby}, Suzanne H. and {Jain}, Bhuvnesh and {Jammes}, Fabrice and {Jee}, M. James and {Jenness}, Tim and {Jernigan}, Garrett and {Jevremovi{\'c}}, Darko and {Johns}, Kenneth and {Johnson}, Anthony S. and {Johnson}, Margaret W.~G. and {Jones}, R. Lynne and {Juramy-Gilles}, Claire and {Juri{\'c}}, Mario and {Kalirai}, Jason S. and {Kallivayalil}, Nitya J. and {Kalmbach}, Bryce and {Kantor}, Jeffrey P. and {Karst}, Pierre and {Kasliwal}, Mansi M. and {Kelly}, Heather and {Kessler}, Richard and {Kinnison}, Veronica and {Kirkby}, David and {Knox}, Lloyd and {Kotov}, Ivan V. and {Krabbendam}, Victor L. and {Krughoff}, K. Simon and {Kub{\'a}nek}, Petr and {Kuczewski}, John and {Kulkarni}, Shri and {Ku}, John and {Kurita}, Nadine R. and {Lage}, Craig S. and {Lambert}, Ron and {Lange}, Travis and {Langton}, J. Brian and {Le Guillou}, Laurent and {Levine}, Deborah and {Liang}, Ming and {Lim}, Kian-Tat and {Lintott}, Chris J. and {Long}, Kevin E. and {Lopez}, Margaux and {Lotz}, Paul J. and {Lupton}, Robert H. and {Lust}, Nate B. and {MacArthur}, Lauren A. and {Mahabal}, Ashish and {Mandelbaum}, Rachel and {Markiewicz}, Thomas W. and {Marsh}, Darren S. and {Marshall}, Philip J. and {Marshall}, Stuart and {May}, Morgan and {McKercher}, Robert and {McQueen}, Michelle and {Meyers}, Joshua and {Migliore}, Myriam and {Miller}, Michelle and {Mills}, David J.},
        title = "{LSST: From Science Drivers to Reference Design and Anticipated Data Products}",
      journal = {\apj},
         year = 2019,
        month = mar,
       volume = {873},
       number = {2},
          eid = {111},
        pages = {111},
          doi = {10.3847/1538-4357/ab042c},
archivePrefix = {arXiv},
       eprint = {0805.2366},
 primaryClass = {astro-ph},
       adsurl = {https://ui.adsabs.harvard.edu/abs/2019ApJ...873..111I}
}

@article{scikit-learn,
 title={Scikit-learn: Machine Learning in {P}ython},
 author={Pedregosa, F. and Varoquaux, G. and Gramfort, A. and Michel, V.
         and Thirion, B. and Grisel, O. and Blondel, M. and Prettenhofer, P.
         and Weiss, R. and Dubourg, V. and Vanderplas, J. and Passos, A. and
         Cournapeau, D. and Brucher, M. and Perrot, M. and Duchesnay, E.},
 journal={Journal of Machine Learning Research},
 volume={12},
 pages={2825--2830},
 year={2011}
}

@ARTICLE{Fadely-2012,
       author = {{Fadely}, Ross and {Hogg}, David W. and {Willman}, Beth},
        title = "{Star-Galaxy Classification in Multi-band Optical Imaging}",
      journal = {\apj},
         year = 2012,
        month = nov,
       volume = {760},
       number = {1},
          eid = {15},
        pages = {15},
          doi = {10.1088/0004-637X/760/1/15},
archivePrefix = {arXiv},
       eprint = {1206.4306},
 primaryClass = {astro-ph.IM},
       adsurl = {https://ui.adsabs.harvard.edu/abs/2012ApJ...760...15F}
}

@misc{NSF-DOE-VRO-2025a,
    author = "{NSF-DOE Vera C. Rubin Observatory}",
    doi = "10.71929/RUBIN/2570325",
    url = "https://www.osti.gov//servlets/purl/2570325",
    title = "{Legacy Survey of Space and Time Data Preview 1: Object searchable catalog [Data set]}",
    publisher = "NSF-DOE Vera C. Rubin Observatory",
    year = "2025"
}

@misc{SF-DOE-VRO-2025b,
    author = "{NSF-DOE Vera C.\ Rubin Observatory}",
    doi = "10.71929/RUBIN/2570308",
    url = "https://www.osti.gov//servlets/purl/2570308",
    title = "{Legacy Survey of Space and Time Data Preview 1 [Data set]}",
    publisher = "NSF-DOE Vera C. Rubin Observatory",
    year = "2025"
}

@ARTICLE{Chambers-2016,
       author = {{Chambers}, K.~C. and {Magnier}, E.~A. and {Metcalfe}, N. and {Flewelling}, H.~A. and {Huber}, M.~E. and {Waters}, C.~Z. and {Denneau}, L. and {Draper}, P.~W. and {Farrow}, D. and {Finkbeiner}, D.~P. and {Holmberg}, C. and {Koppenhoefer}, J. and {Price}, P.~A. and {Rest}, A. and {Saglia}, R.~P. and {Schlafly}, E.~F. and {Smartt}, S.~J. and {Sweeney}, W. and {Wainscoat}, R.~J. and {Burgett}, W.~S. and {Chastel}, S. and {Grav}, T. and {Heasley}, J.~N. and {Hodapp}, K.~W. and {Jedicke}, R. and {Kaiser}, N. and {Kudritzki}, R.-P. and {Luppino}, G.~A. and {Lupton}, R.~H. and {Monet}, D.~G. and {Morgan}, J.~S. and {Onaka}, P.~M. and {Shiao}, B. and {Stubbs}, C.~W. and {Tonry}, J.~L. and {White}, R. and {Ba{\~n}ados}, E. and {Bell}, E.~F. and {Bender}, R. and {Bernard}, E.~J. and {Boegner}, M. and {Boffi}, F. and {Botticella}, M.~T. and {Calamida}, A. and {Casertano}, S. and {Chen}, W.-P. and {Chen}, X. and {Cole}, S. and {Deacon}, N. and {Frenk}, C. and {Fitzsimmons}, A. and {Gezari}, S. and {Gibbs}, V. and {Goessl}, C. and {Goggia}, T. and {Gourgue}, R. and {Goldman}, B. and {Grant}, P. and {Grebel}, E.~K. and {Hambly}, N.~C. and {Hasinger}, G. and {Heavens}, A.~F. and {Heckman}, T.~M. and {Henderson}, R. and {Henning}, T. and {Holman}, M. and {Hopp}, U. and {Ip}, W.-H. and {Isani}, S. and {Jackson}, M. and {Keyes}, C.~D. and {Koekemoer}, A.~M. and {Kotak}, R. and {Le}, D. and {Liska}, D. and {Long}, K.~S. and {Lucey}, J.~R. and {Liu}, M. and {Martin}, N.~F. and {Masci}, G. and {McLean}, B. and {Mindel}, E. and {Misra}, P. and {Morganson}, E. and {Murphy}, D.~N.~A. and {Obaika}, A. and {Narayan}, G. and {Nieto-Santisteban}, M.~A. and {Norberg}, P. and {Peacock}, J.~A. and {Pier}, E.~A. and {Postman}, M. and {Primak}, N. and {Rae}, C. and {Rai}, A. and {Riess}, A. and {Riffeser}, A. and {Rix}, H.~W. and {R{\"o}ser}, S. and {Russel}, R. and {Rutz}, L. and {Schilbach}, E. and {Schultz}, A.~S.~B. and {Scolnic}, D. and {Strolger}, L. and {Szalay}, A. and {Seitz}, S. and {Small}, E. and {Smith}, K.~W. and {Soderblom}, D.~R. and {Taylor}, P. and {Thomson}, R. and {Taylor}, A.~N. and {Thakar}, A.~R. and {Thiel}, J. and {Thilker}, D. and {Unger}, D. and {Urata}, Y. and {Valenti}, J. and {Wagner}, J. and {Walder}, T. and {Walter}, F. and {Watters}, S.~P. and {Werner}, S. and {Wood-Vasey}, W.~M. and {Wyse}, R.},
        title = "{The Pan-STARRS1 Surveys}",
      journal = {arXiv e-prints},
         year = 2016,
        month = dec,
          eid = {arXiv:1612.05560},
        pages = {arXiv:1612.05560},
          doi = {10.48550/arXiv.1612.05560},
archivePrefix = {arXiv},
       eprint = {1612.05560},
 primaryClass = {astro-ph.IM},
       adsurl = {https://ui.adsabs.harvard.edu/abs/2016arXiv161205560C}
}

@ARTICLE{Smolcic-2004,
       author = {{Smol{\v{c}}i{\'c}}, Vernesa and {Ivezi{\'c}}, {\v{Z}}eljko and {Knapp}, Gillian R. and {Lupton}, Robert H. and {Pavlovski}, Kre{\v{s}}imir and {Iliji{\'c}}, Sa{\v{s}}a and {Schlegel}, David and {Smith}, J. Allyn and {McGehee}, Peregrine M. and {Silvestri}, Nicole M. and {Hawley}, Suzanne L. and {Rockosi}, Constance and {Gunn}, James E. and {Strauss}, Michael A. and {Fan}, Xiaohui and {Eisenstein}, Daniel and {Harris}, Hugh},
        title = "{A Second Stellar Color Locus: a Bridge from White Dwarfs to M stars}",
      journal = {\apjl},
         year = 2004,
        month = nov,
       volume = {615},
       number = {2},
        pages = {L141-L144},
          doi = {10.1086/426475},
archivePrefix = {arXiv},
       eprint = {astro-ph/0403218},
 primaryClass = {astro-ph},
       adsurl = {https://ui.adsabs.harvard.edu/abs/2004ApJ...615L.141S}
}

@ARTICLE{Wright-2024,
       author = {{Wright}, Angus H. and {Kuijken}, Konrad and {Hildebrandt}, Hendrik and {Radovich}, Mario and {Bilicki}, Maciej and {Dvornik}, Andrej and {Getman}, Fedor and {Heymans}, Catherine and {Hoekstra}, Henk and {Li}, Shun-Sheng and {Miller}, Lance and {Napolitano}, Nicola R. and {Xia}, Qianli and {Asgari}, Marika and {Brescia}, Massimo and {Buddelmeijer}, Hugo and {Burger}, Pierre and {Castignani}, Gianluca and {Cavuoti}, Stefano and {de Jong}, Jelte and {Edge}, Alastair and {Giblin}, Benjamin and {Giocoli}, Carlo and {Harnois-D{\'e}raps}, Joachim and {Jalan}, Priyanka and {Joachimi}, Benjamin and {John William}, Anjitha and {Joudaki}, Shahab and {Kannawadi}, Arun and {Kaur}, Gursharanjit and {La Barbera}, Francesco and {Linke}, Laila and {Mahony}, Constance and {Maturi}, Matteo and {Moscardini}, Lauro and {Nakoneczny}, Szymon J. and {Paolillo}, Maurizio and {Porth}, Lucas and {Puddu}, Emanuella and {Reischke}, Robert and {Schneider}, Peter and {Sereno}, Mauro and {Shan}, HuanYuan and {Sif{\'o}n}, Crist{\'o}bal and {St{\"o}lzner}, Benjamin and {Tr{\"o}ster}, Tilman and {Valentijn}, Edwin and {van den Busch}, Jan Luca and {Verdoes Kleijn}, Gijs and {Wittje}, Anna and {Yan}, Ziang and {Yao}, Ji and {Yoon}, Mijin and {Zhang}, Yun-Hao},
        title = "{The fifth data release of the Kilo Degree Survey: Multi-epoch optical/NIR imaging covering wide and legacy-calibration fields}",
      journal = {\aap},
         year = 2024,
        month = jun,
       volume = {686},
          eid = {A170},
        pages = {A170},
          doi = {10.1051/0004-6361/202346730},
archivePrefix = {arXiv},
       eprint = {2503.19439},
 primaryClass = {astro-ph.GA},
       adsurl = {https://ui.adsabs.harvard.edu/abs/2024A&A...686A.170W}
}

@ARTICLE{Feng-2025,
       author = {{Feng}, Hai-Cheng and {Li}, Rui and {Napolitano}, Nicola R. and {Li}, Sha-Sha and {Bai}, J.~M. and {Dong}, Yue and {Li}, Ran and {Liu}, H.~T. and {Lu}, Kai-Xing and {Pan}, Zhi-Wei and {Radovich}, Mario and {Shan}, Huan-Yuan and {Wang}, Jian-Guo and {Xi}, Wen-Zhe and {Xie}, Ling-Hua and {Yuan}, Zun-Li and {Zhang}, Yang-Wei},
        title = "{Morpho-photometric Classification of KiDS DR5 Sources Based on Neural Networks: A Comprehensive Star{\textendash}Quasar{\textendash}Galaxy Catalog}",
      journal = {\apjs},
         year = 2025,
        month = jul,
       volume = {279},
       number = {1},
          eid = {26},
        pages = {26},
          doi = {10.3847/1538-4365/adde5a},
archivePrefix = {arXiv},
       eprint = {2406.03797},
 primaryClass = {astro-ph.GA},
       adsurl = {https://ui.adsabs.harvard.edu/abs/2025ApJS..279...26F}
}

@ARTICLE{Scoville-2007,
       author = {{Scoville}, N. and {Aussel}, H. and {Brusa}, M. and {Capak}, P. and {Carollo}, C.~M. and {Elvis}, M. and {Giavalisco}, M. and {Guzzo}, L. and {Hasinger}, G. and {Impey}, C. and {Kneib}, J. -P. and {LeFevre}, O. and {Lilly}, S.~J. and {Mobasher}, B. and {Renzini}, A. and {Rich}, R.~M. and {Sanders}, D.~B. and {Schinnerer}, E. and {Schminovich}, D. and {Shopbell}, P. and {Taniguchi}, Y. and {Tyson}, N.~D.},
        title = "{The Cosmic Evolution Survey (COSMOS): Overview}",
      journal = {\apjs},
         year = 2007,
        month = sep,
       volume = {172},
       number = {1},
        pages = {1-8},
          doi = {10.1086/516585},
archivePrefix = {arXiv},
       eprint = {astro-ph/0612305},
 primaryClass = {astro-ph},
       adsurl = {https://ui.adsabs.harvard.edu/abs/2007ApJS..172....1S}
}

@ARTICLE{Choi-2025,
       author = {{Choi}, Yumi and {Olsen}, Knut A.~G. and {Carlin}, Jeffrey L. and {Yuankun} and {Wang} and {Moolekamp}, Fred and {Saha}, Abi and {Sullivan}, Ian and {Slater}, Colin T. and {Tucker}, Douglas L. and {Adair}, Christina L. and {Ferguson}, Peter S. and {Kang}, Yijung and {Pe{\~n}a Ram{\'\i}rez}, Karla and {Rabus}, Markus},
        title = "{47 Tuc in Rubin Data Preview 1: Exploring Early LSST Data and Science Potential}",
      journal = {arXiv e-prints},
         year = 2025,
        month = jul,
          eid = {arXiv:2507.01343},
        pages = {arXiv:2507.01343},
          doi = {10.48550/arXiv.2507.01343},
archivePrefix = {arXiv},
       eprint = {2507.01343},
 primaryClass = {astro-ph.SR},
       adsurl = {https://ui.adsabs.harvard.edu/abs/2025arXiv250701343C}
}

@TechReport{VRO-DP1,
    author = "{NSF-DOE Vera C. Rubin Observatory}",
    title = "{The Vera C. Rubin Observatory Data Preview 1}",
    institution = "{NSF-DOE Vera C. Rubin Observatory}",
    year = "2025",
    month = "July",
    handle = "RTN-095",
    type = "{Rubin Technical Note}",
    number = "RTN-095",
    doi = "10.71929/rubin/2570536",
    url = "https://rtn-095.lsst.io/"
}

@ARTICLE{Skelton-2014,
       author = {{Skelton}, Rosalind E. and {Whitaker}, Katherine E. and {Momcheva}, Ivelina G. and {Brammer}, Gabriel B. and {van Dokkum}, Pieter G. and {Labb{\'e}}, Ivo and {Franx}, Marijn and {van der Wel}, Arjen and {Bezanson}, Rachel and {Da Cunha}, Elisabete and {Fumagalli}, Mattia and {F{\"o}rster Schreiber}, Natascha and {Kriek}, Mariska and {Leja}, Joel and {Lundgren}, Britt F. and {Magee}, Daniel and {Marchesini}, Danilo and {Maseda}, Michael V. and {Nelson}, Erica J. and {Oesch}, Pascal and {Pacifici}, Camilla and {Patel}, Shannon G. and {Price}, Sedona and {Rix}, Hans-Walter and {Tal}, Tomer and {Wake}, David A. and {Wuyts}, Stijn},
        title = "{3D-HST WFC3-selected Photometric Catalogs in the Five CANDELS/3D-HST Fields: Photometry, Photometric Redshifts, and Stellar Masses}",
      journal = {\apjs},
         year = 2014,
        month = oct,
       volume = {214},
       number = {2},
          eid = {24},
        pages = {24},
          doi = {10.1088/0067-0049/214/2/24},
archivePrefix = {arXiv},
       eprint = {1403.3689},
 primaryClass = {astro-ph.GA},
       adsurl = {https://ui.adsabs.harvard.edu/abs/2014ApJS..214...24S}
}

@ARTICLE{Maddox-2008,
       author = {{Maddox}, Natasha and {Hewett}, Paul C. and {Warren}, S.~J. and {Croom}, S.~M.},
        title = "{Luminous K-band selected quasars from UKIDSS}",
      journal = {\mnras},
         year = 2008,
        month = may,
       volume = {386},
       number = {3},
        pages = {1605-1624},
          doi = {10.1111/j.1365-2966.2008.13138.x},
archivePrefix = {arXiv},
       eprint = {0802.3650},
 primaryClass = {astro-ph},
       adsurl = {https://ui.adsabs.harvard.edu/abs/2008MNRAS.386.1605M}
}

@software{topcat,
       author = {{Taylor}, Mark},
        title = "{TOPCAT: Tool for OPerations on Catalogues And Tables}",
 howpublished = {Astrophysics Source Code Library, record ascl:1101.010},
         year = 2011,
        month = jan,
          eid = {ascl:1101.010},
archivePrefix = {ascl},
       eprint = {1101.010},
       adsurl = {https://ui.adsabs.harvard.edu/abs/2011ascl.soft01010T}
}

@Article{numpy,
 title         = {Array programming with {NumPy}},
 author        = {Charles R. Harris and K. Jarrod Millman and St{\'{e}}fan J.
                 van der Walt and Ralf Gommers and Pauli Virtanen and David
                 Cournapeau and Eric Wieser and Julian Taylor and Sebastian
                 Berg and Nathaniel J. Smith and Robert Kern and Matti Picus
                 and Stephan Hoyer and Marten H. van Kerkwijk and Matthew
                 Brett and Allan Haldane and Jaime Fern{\'{a}}ndez del
                 R{\'{i}}o and Mark Wiebe and Pearu Peterson and Pierre
                 G{\'{e}}rard-Marchant and Kevin Sheppard and Tyler Reddy and
                 Warren Weckesser and Hameer Abbasi and Christoph Gohlke and
                 Travis E. Oliphant},
 year          = {2020},
 month         = sep,
 journal       = {Nature},
 volume        = {585},
 number        = {7825},
 pages         = {357--362},
 doi           = {10.1038/s41586-020-2649-2},
 publisher     = {Springer Science and Business Media {LLC}},
 url           = {https://doi.org/10.1038/s41586-020-2649-2}
}

@software{pandas-2020,
    author       = {The pandas development team},
    title        = {pandas-dev/pandas: Pandas},
    month        = feb,
    year         = 2020,
    publisher    = {Zenodo},
    version      = {latest},
    doi          = {10.5281/zenodo.3509134},
    url          = {https://doi.org/10.5281/zenodo.3509134}
}

@Manual{XGBoost-2026,
  title = {xgboost: Extreme Gradient Boosting},
  author = {Tianqi Chen and Tong He and Michael Benesty and Vadim Khotilovich and Yuan Tang and Hyunsu Cho and Kailong Chen and Rory Mitchell and Ignacio Cano and Tianyi Zhou and Mu Li and Junyuan Xie and Min Lin and Yifeng Geng and Yutian Li and Jiaming Yuan and David Cortes},
  year = {2026},
  note = {R package version 3.3.0.0},
  url = {https://github.com/dmlc/xgboost},
}

@inproceedings{XGBoost-2016,
 author = {Chen, Tianqi and Guestrin, Carlos},
 title = {{XGBoost}: A Scalable Tree Boosting System},
 booktitle = {Proceedings of the 22nd ACM SIGKDD International Conference on Knowledge Discovery and Data Mining},
 series = {KDD '16},
 year = {2016},
 isbn = {978-1-4503-4232-2},
 location = {San Francisco, California, USA},
 pages = {785--794},
 numpages = {10},
 url = {http://doi.acm.org/10.1145/2939672.2939785},
 doi = {10.1145/2939672.2939785},
 acmid = {2939785},
 publisher = {ACM},
 address = {New York, NY, USA},
}

@InProceedings{pandas-2010,
  author    = { {W}es {M}c{K}inney },
  title     = { {D}ata {S}tructures for {S}tatistical {C}omputing in {P}ython },
  booktitle = { {P}roceedings of the 9th {P}ython in {S}cience {C}onference },
  pages     = { 56 - 61 },
  year      = { 2010 },
  editor    = { {S}t\'efan van der {W}alt and {J}arrod {M}illman },
  doi       = { 10.25080/Majora-92bf1922-00a }
}

@Article{matplotlib,
  Author    = {Hunter, J. D.},
  Title     = {Matplotlib: A 2D graphics environment},
  Journal   = {Computing in Science \& Engineering},
  Volume    = {9},
  Number    = {3},
  Pages     = {90--95},
  publisher = {IEEE COMPUTER SOC},
  doi       = {10.1109/MCSE.2007.55},
  year      = 2007
}

@ARTICLE{Tortora-2018,
       author = {{Tortora}, C. and {Napolitano}, N.~R. and {Spavone}, M. and {La Barbera}, F. and {D'Ago}, G. and {Spiniello}, C. and {Kuijken}, K.~H. and {Roy}, N. and {Raj}, M.~A. and {Cavuoti}, S. and {Brescia}, M. and {Longo}, G. and {Pota}, V. and {Petrillo}, C.~E. and {Radovich}, M. and {Getman}, F. and {Koopmans}, L.~V.~E. and {Trujillo}, I. and {Verdoes Kleijn}, G. and {Capaccioli}, M. and {Grado}, A. and {Covone}, G. and {Scognamiglio}, D. and {Blake}, C. and {Glazebrook}, K. and {Joudaki}, S. and {Lidman}, C. and {Wolf}, C.},
        title = "{The first sample of spectroscopically confirmed ultra-compact massive galaxies in the Kilo Degree Survey}",
      journal = {\mnras},
         year = 2018,
        month = dec,
       volume = {481},
       number = {4},
        pages = {4728-4752},
          doi = {10.1093/mnras/sty2564},
archivePrefix = {arXiv},
       eprint = {1806.01307},
 primaryClass = {astro-ph.GA},
       adsurl = {https://ui.adsabs.harvard.edu/abs/2018MNRAS.481.4728T}
}

@ARTICLE{Khramtsov-2019,
       author = {{Khramtsov}, Vladislav and {Sergeyev}, Alexey and {Spiniello}, Chiara and {Tortora}, Crescenzo and {Napolitano}, Nicola R. and {Agnello}, Adriano and {Getman}, Fedor and {de Jong}, Jelte T.~A. and {Kuijken}, Konrad and {Radovich}, Mario and {Shan}, HuanYuan and {Shulga}, Valery},
        title = "{KiDS-SQuaD. II. Machine learning selection of bright extragalactic objects to search for new gravitationally lensed quasars}",
      journal = {\aap},
         year = 2019,
        month = dec,
       volume = {632},
          eid = {A56},
        pages = {A56},
          doi = {10.1051/0004-6361/201936006},
archivePrefix = {arXiv},
       eprint = {1906.01638},
 primaryClass = {astro-ph.GA},
       adsurl = {https://ui.adsabs.harvard.edu/abs/2019A&A...632A..56K}
}

@ARTICLE{Jeakel-2026,
       author = {{Jeakel}, Ana Paula and {Vieira dos Santos}, Gabriel and {Marra}, Valerio and {von Marttens}, Rodrigo and {Gurung-L{\'o}pez}, Siddhartha and {Abramo}, Raul and {Alcaniz}, Jailson and {Benitez}, Narciso and {Bonoli}, Silvia and {Cenarro}, Javier and {Crist{\'o}bal-Hornillos}, David and {Daflon}, Simone and {Dupke}, Renato and {Ederoclite}, Alessandro and {Gonz{\'a}lez Delgado}, Rosa M. and {Hern{\'a}n-Caballero}, Antonio and {Hern{\'a}ndez-Monteagudo}, Carlos and {Liu}, Jifeng and {L{\'o}pez-Sanjuan}, Carlos and {Mar{\'\i}n-Franch}, Antonio and {Mendes de Oliveira}, Claudia and {Moles}, Mariano and {Roig}, Fernando and {Sodr{\'e}}, Jr., Laerte and {Taylor}, Keith and {Varela}, Jes{\'u}s and {V{\'a}zquez Rami{\'o}}, H{\'e}ctor and {Vilchez}, Jos{\'e} M. and {Willmer}, Christopher and {Zaragoza-Cardiel}, Javier},
        title = "{The miniJPAS and J-NEP Surveys: Machine Learning for Star-Galaxy Separation}",
      journal = {Galaxies},
         year = 2026,
        month = jan,
       volume = {14},
       number = {1},
          eid = {6},
        pages = {6},
          doi = {10.3390/galaxies14010006},
archivePrefix = {arXiv},
       eprint = {2511.20524},
 primaryClass = {astro-ph.IM},
       adsurl = {https://ui.adsabs.harvard.edu/abs/2026Galax..14....6J}
}

@ARTICLE{Kurk-2013,
       author = {{Kurk}, J. and {Cimatti}, A. and {Daddi}, E. and {Mignoli}, M. and {Pozzetti}, L. and {Dickinson}, M. and {Bolzonella}, M. and {Zamorani}, G. and {Cassata}, P. and {Rodighiero}, G. and {Franceschini}, A. and {Renzini}, A. and {Rosati}, P. and {Halliday}, C. and {Berta}, S.},
        title = "{GMASS ultradeep spectroscopy of galaxies at z \raisebox{-0.5ex}\textasciitilde 2. VII. Sample selection and spectroscopy}",
      journal = {\aap},
         year = 2013,
        month = jan,
       volume = {549},
          eid = {A63},
        pages = {A63},
          doi = {10.1051/0004-6361/201117847},
archivePrefix = {arXiv},
       eprint = {1209.1561},
 primaryClass = {astro-ph.CO},
       adsurl = {https://ui.adsabs.harvard.edu/abs/2013A&A...549A..63K}
}

@ARTICLE{Walsh-2009,
       author = {{Walsh}, S.~M. and {Willman}, B. and {Jerjen}, H.},
        title = "{The Invisibles: A Detection Algorithm to Trace the Faintest Milky Way Satellites}",
      journal = {\aj},
         year = 2009,
        month = jan,
       volume = {137},
       number = {1},
        pages = {450-469},
          doi = {10.1088/0004-6256/137/1/450},
archivePrefix = {arXiv},
       eprint = {0807.3345},
 primaryClass = {astro-ph},
       adsurl = {https://ui.adsabs.harvard.edu/abs/2009AJ....137..450W}
}

@ARTICLE{Schlafly&Finkbeiner2011,
       author = {{Schlafly}, Edward F. and {Finkbeiner}, Douglas P.},
        title = "{Measuring Reddening with Sloan Digital Sky Survey Stellar Spectra and Recalibrating SFD}",
      journal = {\apj},
         year = "2011",
        month = "Aug",
       volume = {737},
       number = {2},
          eid = {103},
        pages = {103},
          doi = {10.1088/0004-637X/737/2/103},
archivePrefix = {arXiv},
       eprint = {1012.4804},
 primaryClass = {astro-ph.GA},
       adsurl = {https://ui.adsabs.harvard.edu/abs/2011ApJ...737..103S}
}

@ARTICLE{Gaia-Prusti-2016,
       author = {{Gaia Collaboration} and {Prusti}, T. and {de Bruijne}, J.~H.~J. and {Brown}, A.~G.~A. and {Vallenari}, A. and {Babusiaux}, C. and {Bailer-Jones}, C.~A.~L. and {Bastian}, U. and {Biermann}, M. and {Evans}, D.~W. and {Eyer}, L. and {Jansen}, F. and {Jordi}, C. and {Klioner}, S.~A. and {Lammers}, U. and {Lindegren}, L. and {Luri}, X. and {Mignard}, F. and {Milligan}, D.~J. and {Panem}, C. and {Poinsignon}, V. and {Pourbaix}, D. and {Randich}, S. and {Sarri}, G. and {Sartoretti}, P. and {Siddiqui}, H.~I. and {Soubiran}, C. and {Valette}, V. and {van Leeuwen}, F. and {Walton}, N.~A. and {Aerts}, C. and {Arenou}, F. and {Cropper}, M. and {Drimmel}, R. and {H{\o}g}, E. and {Katz}, D. and {Lattanzi}, M.~G. and {O'Mullane}, W. and {Grebel}, E.~K. and {Holland}, A.~D. and {Huc}, C. and {Passot}, X. and {Bramante}, L. and {Cacciari}, C. and {Casta{\~n}eda}, J. and {Chaoul}, L. and {Cheek}, N. and {De Angeli}, F. and {Fabricius}, C. and {Guerra}, R. and {Hern{\'a}ndez}, J. and {Jean-Antoine-Piccolo}, A. and {Masana}, E. and {Messineo}, R. and {Mowlavi}, N. and {Nienartowicz}, K. and {Ord{\'o}{\~n}ez-Blanco}, D. and {Panuzzo}, P. and {Portell}, J. and {Richards}, P.~J. and {Riello}, M. and {Seabroke}, G.~M. and {Tanga}, P. and {Th{\'e}venin}, F. and {Torra}, J. and {Els}, S.~G. and {Gracia-Abril}, G. and {Comoretto}, G. and {Garcia-Reinaldos}, M. and {Lock}, T. and {Mercier}, E. and {Altmann}, M. and {Andrae}, R. and {Astraatmadja}, T.~L. and {Bellas-Velidis}, I. and {Benson}, K. and {Berthier}, J. and {Blomme}, R. and {Busso}, G. and {Carry}, B. and {Cellino}, A. and {Clementini}, G. and {Cowell}, S. and {Creevey}, O. and {Cuypers}, J. and {Davidson}, M. and {De Ridder}, J. and {de Torres}, A. and {Delchambre}, L. and {Dell'Oro}, A. and {Ducourant}, C. and {Fr{\'e}mat}, Y. and {Garc{\'\i}a-Torres}, M. and {Gosset}, E. and {Halbwachs}, J. -L. and {Hambly}, N.~C. and {Harrison}, D.~L. and {Hauser}, M. and {Hestroffer}, D. and {Hodgkin}, S.~T. and {Huckle}, H.~E. and {Hutton}, A. and {Jasniewicz}, G. and {Jordan}, S. and {Kontizas}, M. and {Korn}, A.~J. and {Lanzafame}, A.~C. and {Manteiga}, M. and {Moitinho}, A. and {Muinonen}, K. and {Osinde}, J. and {Pancino}, E. and {Pauwels}, T. and {Petit}, J. -M. and {Recio-Blanco}, A. and {Robin}, A.~C. and {Sarro}, L.~M. and {Siopis}, C. and {Smith}, M. and {Smith}, K.~W. and {Sozzetti}, A. and {Thuillot}, W. and {van Reeven}, W. and {Viala}, Y. and {Abbas}, U. and {Abreu Aramburu}, A. and {Accart}, S. and {Aguado}, J.~J. and {Allan}, P.~M. and {Allasia}, W. and {Altavilla}, G. and {{\'A}lvarez}, M.~A. and {Alves}, J. and {Anderson}, R.~I. and {Andrei}, A.~H. and {Anglada Varela}, E. and {Antiche}, E. and {Antoja}, T. and {Ant{\'o}n}, S. and {Arcay}, B. and {Atzei}, A. and {Ayache}, L. and {Bach}, N. and {Baker}, S.~G. and {Balaguer-N{\'u}{\~n}ez}, L. and {Barache}, C. and {Barata}, C. and {Barbier}, A. and {Barblan}, F. and {Baroni}, M. and {Barrado y Navascu{\'e}s}, D. and {Barros}, M. and {Barstow}, M.~A. and {Becciani}, U. and {Bellazzini}, M. and {Bellei}, G. and {Bello Garc{\'\i}a}, A. and {Belokurov}, V. and {Bendjoya}, P. and {Berihuete}, A. and {Bianchi}, L. and {Bienaym{\'e}}, O. and {Billebaud}, F. and {Blagorodnova}, N. and {Blanco-Cuaresma}, S. and {Boch}, T. and {Bombrun}, A. and {Borrachero}, R. and {Bouquillon}, S. and {Bourda}, G. and {Bouy}, H. and {Bragaglia}, A. and {Breddels}, M.~A. and {Brouillet}, N. and {Br{\"u}semeister}, T. and {Bucciarelli}, B. and {Budnik}, F. and {Burgess}, P. and {Burgon}, R. and {Burlacu}, A. and {Busonero}, D. and {Buzzi}, R. and {Caffau}, E. and {Cambras}, J. and {Campbell}, H. and {Cancelliere}, R. and {Cantat-Gaudin}, T. and {Carlucci}, T. and {Carrasco}, J.~M. and {Castellani}, M. and {Charlot}, P. and {Charnas}, J. and {Charvet}, P. and {Chassat}, F. and {Chiavassa}, A. and {Clotet}, M. and {Cocozza}, G. and {Collins}, R.~S. and {Collins}, P. and {Costigan}, G.},
        title = "{The Gaia mission}",
      journal = {\aap},
         year = 2016,
        month = nov,
       volume = {595},
          eid = {A1},
        pages = {A1},
          doi = {10.1051/0004-6361/201629272},
archivePrefix = {arXiv},
       eprint = {1609.04153},
 primaryClass = {astro-ph.IM},
       adsurl = {https://ui.adsabs.harvard.edu/abs/2016A&A...595A...1G}
}

@ARTICLE{Gaia-Vallenari-2023,
       author = {{Gaia Collaboration} and {Vallenari}, A. and {Brown}, A.~G.~A. and {Prusti}, T. and {de Bruijne}, J.~H.~J. and {Arenou}, F. and {Babusiaux}, C. and {Biermann}, M. and {Creevey}, O.~L. and {Ducourant}, C. and {Evans}, D.~W. and {Eyer}, L. and {Guerra}, R. and {Hutton}, A. and {Jordi}, C. and {Klioner}, S.~A. and {Lammers}, U.~L. and {Lindegren}, L. and {Luri}, X. and {Mignard}, F. and {Panem}, C. and {Pourbaix}, D. and {Randich}, S. and {Sartoretti}, P. and {Soubiran}, C. and {Tanga}, P. and {Walton}, N.~A. and {Bailer-Jones}, C.~A.~L. and {Bastian}, U. and {Drimmel}, R. and {Jansen}, F. and {Katz}, D. and {Lattanzi}, M.~G. and {van Leeuwen}, F. and {Bakker}, J. and {Cacciari}, C. and {Casta{\~n}eda}, J. and {De Angeli}, F. and {Fabricius}, C. and {Fouesneau}, M. and {Fr{\'e}mat}, Y. and {Galluccio}, L. and {Guerrier}, A. and {Heiter}, U. and {Masana}, E. and {Messineo}, R. and {Mowlavi}, N. and {Nicolas}, C. and {Nienartowicz}, K. and {Pailler}, F. and {Panuzzo}, P. and {Riclet}, F. and {Roux}, W. and {Seabroke}, G.~M. and {Sordo}, R. and {Th{\'e}venin}, F. and {Gracia-Abril}, G. and {Portell}, J. and {Teyssier}, D. and {Altmann}, M. and {Andrae}, R. and {Audard}, M. and {Bellas-Velidis}, I. and {Benson}, K. and {Berthier}, J. and {Blomme}, R. and {Burgess}, P.~W. and {Busonero}, D. and {Busso}, G. and {C{\'a}novas}, H. and {Carry}, B. and {Cellino}, A. and {Cheek}, N. and {Clementini}, G. and {Damerdji}, Y. and {Davidson}, M. and {de Teodoro}, P. and {Nu{\~n}ez Campos}, M. and {Delchambre}, L. and {Dell'Oro}, A. and {Esquej}, P. and {Fern{\'a}ndez-Hern{\'a}ndez}, J. and {Fraile}, E. and {Garabato}, D. and {Garc{\'\i}a-Lario}, P. and {Gosset}, E. and {Haigron}, R. and {Halbwachs}, J. -L. and {Hambly}, N.~C. and {Harrison}, D.~L. and {Hern{\'a}ndez}, J. and {Hestroffer}, D. and {Hodgkin}, S.~T. and {Holl}, B. and {Jan{\ss}en}, K. and {Jevardat de Fombelle}, G. and {Jordan}, S. and {Krone-Martins}, A. and {Lanzafame}, A.~C. and {L{\"o}ffler}, W. and {Marchal}, O. and {Marrese}, P.~M. and {Moitinho}, A. and {Muinonen}, K. and {Osborne}, P. and {Pancino}, E. and {Pauwels}, T. and {Recio-Blanco}, A. and {Reyl{\'e}}, C. and {Riello}, M. and {Rimoldini}, L. and {Roegiers}, T. and {Rybizki}, J. and {Sarro}, L.~M. and {Siopis}, C. and {Smith}, M. and {Sozzetti}, A. and {Utrilla}, E. and {van Leeuwen}, M. and {Abbas}, U. and {{\'A}brah{\'a}m}, P. and {Abreu Aramburu}, A. and {Aerts}, C. and {Aguado}, J.~J. and {Ajaj}, M. and {Aldea-Montero}, F. and {Altavilla}, G. and {{\'A}lvarez}, M.~A. and {Alves}, J. and {Anders}, F. and {Anderson}, R.~I. and {Anglada Varela}, E. and {Antoja}, T. and {Baines}, D. and {Baker}, S.~G. and {Balaguer-N{\'u}{\~n}ez}, L. and {Balbinot}, E. and {Balog}, Z. and {Barache}, C. and {Barbato}, D. and {Barros}, M. and {Barstow}, M.~A. and {Bartolom{\'e}}, S. and {Bassilana}, J. -L. and {Bauchet}, N. and {Becciani}, U. and {Bellazzini}, M. and {Berihuete}, A. and {Bernet}, M. and {Bertone}, S. and {Bianchi}, L. and {Binnenfeld}, A. and {Blanco-Cuaresma}, S. and {Blazere}, A. and {Boch}, T. and {Bombrun}, A. and {Bossini}, D. and {Bouquillon}, S. and {Bragaglia}, A. and {Bramante}, L. and {Breedt}, E. and {Bressan}, A. and {Brouillet}, N. and {Brugaletta}, E. and {Bucciarelli}, B. and {Burlacu}, A. and {Butkevich}, A.~G. and {Buzzi}, R. and {Caffau}, E. and {Cancelliere}, R. and {Cantat-Gaudin}, T. and {Carballo}, R. and {Carlucci}, T. and {Carnerero}, M.~I. and {Carrasco}, J.~M. and {Casamiquela}, L. and {Castellani}, M. and {Castro-Ginard}, A. and {Chaoul}, L. and {Charlot}, P. and {Chemin}, L. and {Chiaramida}, V. and {Chiavassa}, A. and {Chornay}, N. and {Comoretto}, G. and {Contursi}, G. and {Cooper}, W.~J. and {Cornez}, T. and {Cowell}, S. and {Crifo}, F. and {Cropper}, M. and {Crosta}, M. and {Crowley}, C. and {Dafonte}, C. and {Dapergolas}, A. and {David}, M. and {David}, P. and {de Laverny}, P. and {De Luise}, F. and {De March}, R.},
        title = "{Gaia Data Release 3. Summary of the content and survey properties}",
      journal = {\aap},
         year = 2023,
        month = jun,
       volume = {674},
          eid = {A1},
        pages = {A1},
          doi = {10.1051/0004-6361/202243940},
archivePrefix = {arXiv},
       eprint = {2208.00211},
 primaryClass = {astro-ph.GA},
       adsurl = {https://ui.adsabs.harvard.edu/abs/2023A&A...674A...1G}
}

@ARTICLE{Logan&Fotopoulou2020,
       author = {{Logan}, C.~H.~A. and {Fotopoulou}, S.},
        title = "{Unsupervised star, galaxy, QSO classification. Application of HDBSCAN}",
      journal = {\aap},
         year = 2020,
        month = jan,
       volume = {633},
          eid = {A154},
        pages = {A154},
          doi = {10.1051/0004-6361/201936648},
archivePrefix = {arXiv},
       eprint = {1911.05107},
 primaryClass = {astro-ph.GA},
       adsurl = {https://ui.adsabs.harvard.edu/abs/2020A&A...633A.154L}
}

@ARTICLE{Zhang-2025,
       author = {{Zhang}, Shiyang and {Hildebrandt}, Hendrik and {Yan}, Ziang and {Smith}, Simon E.~T. and {Gatto}, Massimiliano and {Dall'Ora}, Massimo and {Tortora}, Crescenzo and {Li}, Shun-Sheng and {Els{\"a}sser}, Dominik},
        title = "{Quantifying the detectability of Milky Way satellites with image simulations: Case study with KiDS}",
      journal = {\aap},
         year = 2025,
        month = jun,
       volume = {698},
          eid = {A108},
        pages = {A108},
          doi = {10.1051/0004-6361/202554189},
archivePrefix = {arXiv},
       eprint = {2502.13858},
 primaryClass = {astro-ph.CO},
       adsurl = {https://ui.adsabs.harvard.edu/abs/2025A&A...698A.108Z}
}

@ARTICLE{Fotopoulou&Paltani2018,
       author = {{Fotopoulou}, S. and {Paltani}, S.},
        title = "{CPz: Classification-aided photometric-redshift estimation}",
      journal = {\aap},
         year = 2018,
        month = oct,
       volume = {619},
          eid = {A14},
        pages = {A14},
          doi = {10.1051/0004-6361/201730763},
archivePrefix = {arXiv},
       eprint = {1808.04977},
 primaryClass = {astro-ph.GA},
       adsurl = {https://ui.adsabs.harvard.edu/abs/2018A&A...619A..14F}
}

@ARTICLE{Kodra-2023,
       author = {{Kodra}, Dritan and {Andrews}, Brett H. and {Newman}, Jeffrey A. and {Finkelstein}, Steven L. and {Fontana}, Adriano and {Hathi}, Nimish and {Salvato}, Mara and {Wiklind}, Tommy and {Wuyts}, Stijn and {Broussard}, Adam and {Chartab}, Nima and {Conselice}, Christopher and {Cooper}, M.~C. and {Dekel}, Avishai and {Dickinson}, Mark and {Ferguson}, Henry C. and {Gawiser}, Eric and {Grogin}, Norman A. and {Iyer}, Kartheik and {Kartaltepe}, Jeyhan and {Kassin}, Susan and {Koekemoer}, Anton M. and {Koo}, David C. and {Lucas}, Ray A. and {Mantha}, Kameswara Bharadwaj and {McIntosh}, Daniel H. and {Mobasher}, Bahram and {Pacifici}, Camilla and {P{\'e}rez-Gonz{\'a}lez}, Pablo G. and {Santini}, Paola},
        title = "{Optimized Photometric Redshifts for the Cosmic Assembly Near-infrared Deep Extragalactic Legacy Survey (CANDELS)}",
      journal = {\apj},
         year = 2023,
        month = jan,
       volume = {942},
       number = {1},
          eid = {36},
        pages = {36},
          doi = {10.3847/1538-4357/ac9f12},
archivePrefix = {arXiv},
       eprint = {2210.01140},
 primaryClass = {astro-ph.GA},
       adsurl = {https://ui.adsabs.harvard.edu/abs/2023ApJ...942...36K}
}

@ARTICLE{LeFevre-2013,
       author = {{Le F{\`e}vre}, O. and {Cassata}, P. and {Cucciati}, O. and {Garilli}, B. and {Ilbert}, O. and {Le Brun}, V. and {Maccagni}, D. and {Moreau}, C. and {Scodeggio}, M. and {Tresse}, L. and {Zamorani}, G. and {Adami}, C. and {Arnouts}, S. and {Bardelli}, S. and {Bolzonella}, M. and {Bondi}, M. and {Bongiorno}, A. and {Bottini}, D. and {Cappi}, A. and {Charlot}, S. and {Ciliegi}, P. and {Contini}, T. and {de la Torre}, S. and {Foucaud}, S. and {Franzetti}, P. and {Gavignaud}, I. and {Guzzo}, L. and {Iovino}, A. and {Lemaux}, B. and {L{\'o}pez-Sanjuan}, C. and {McCracken}, H.~J. and {Marano}, B. and {Marinoni}, C. and {Mazure}, A. and {Mellier}, Y. and {Merighi}, R. and {Merluzzi}, P. and {Paltani}, S. and {Pell{\`o}}, R. and {Pollo}, A. and {Pozzetti}, L. and {Scaramella}, R. and {Tasca}, L. and {Vergani}, D. and {Vettolani}, G. and {Zanichelli}, A. and {Zucca}, E.},
        title = "{The VIMOS VLT Deep Survey final data release: a spectroscopic sample of 35 016 galaxies and AGN out to z \raisebox{-0.5ex}\textasciitilde 6.7 selected with 17.5 {\ensuremath{\leq}} i$_{AB}$ {\ensuremath{\leq}} 24.75}",
      journal = {\aap},
         year = 2013,
        month = nov,
       volume = {559},
          eid = {A14},
        pages = {A14},
          doi = {10.1051/0004-6361/201322179},
archivePrefix = {arXiv},
       eprint = {1307.6518},
 primaryClass = {astro-ph.CO},
       adsurl = {https://ui.adsabs.harvard.edu/abs/2013A&A...559A..14L}
}

@ARTICLE{Damen-2011,
       author = {{Damen}, M. and {Labb{\'e}}, I. and {van Dokkum}, P.~G. and {Franx}, M. and {Taylor}, E.~N. and {Brandt}, W.~N. and {Dickinson}, M. and {Gawiser}, E. and {Illingworth}, G.~D. and {Kriek}, M. and {Marchesini}, D. and {Muzzin}, A. and {Papovich}, C. and {Rix}, H.-W.},
        title = "{The SIMPLE Survey: Observations, Reduction, and Catalog}",
      journal = {\apj},
         year = 2011,
        month = jan,
       volume = {727},
       number = {1},
          eid = {1},
        pages = {1},
          doi = {10.1088/0004-637X/727/1/1},
archivePrefix = {arXiv},
       eprint = {1011.2764},
 primaryClass = {astro-ph.CO},
       adsurl = {https://ui.adsabs.harvard.edu/abs/2011ApJ...727....1D}
}

@ARTICLE{randomforest,
       author = {{Breiman}, Leo},
        title = "{Random Forests.}",
      journal = {Machine Learning},
         year = 2001,
        month = jan,
       volume = {45},
        pages = {5-32},
          doi = {10.1023/A:1010933404324},
       adsurl = {https://ui.adsabs.harvard.edu/abs/2001MachL..45....5B}
}

@ARTICLE{Duncan-2022,
       author = {{Duncan}, Kenneth J.},
        title = "{All-purpose, all-sky photometric redshifts for the Legacy Imaging Surveys Data Release 8}",
      journal = {\mnras},
         year = 2022,
        month = may,
       volume = {512},
       number = {3},
        pages = {3662-3683},
          doi = {10.1093/mnras/stac608},
archivePrefix = {arXiv},
       eprint = {2203.01949},
 primaryClass = {astro-ph.GA},
       adsurl = {https://ui.adsabs.harvard.edu/abs/2022MNRAS.512.3662D}
}

@ARTICLE{Wolf-2004,
       author = {{Wolf}, C. and {Meisenheimer}, K. and {Kleinheinrich}, M. and {Borch}, A. and {Dye}, S. and {Gray}, M. and {Wisotzki}, L. and {Bell}, E.~F. and {Rix}, H.-W. and {Cimatti}, A. and {Hasinger}, G. and {Szokoly}, G.},
        title = "{A catalogue of the Chandra Deep Field South with multi-colour classification and photometric redshifts from COMBO-17}",
      journal = {\aap},
         year = 2004,
        month = jul,
       volume = {421},
        pages = {913-936},
          doi = {10.1051/0004-6361:20040525},
archivePrefix = {arXiv},
       eprint = {astro-ph/0403666},
 primaryClass = {astro-ph},
       adsurl = {https://ui.adsabs.harvard.edu/abs/2004A&A...421..913W}
}

@ARTICLE{Cooper-2012,
       author = {{Cooper}, Michael C. and {Yan}, Renbin and {Dickinson}, Mark and {Juneau}, St{\'e}phanie and {Lotz}, Jennifer M. and {Newman}, Jeffrey A. and {Papovich}, Casey and {Salim}, Samir and {Walth}, Gregory and {Weiner}, Benjamin J. and {Willmer}, Christopher N.~A.},
        title = "{The Arizona CDFS Environment Survey (ACES): A Magellan/IMACS Spectroscopic Survey of the Chandra Deep Field-South}",
      journal = {\mnras},
         year = 2012,
        month = sep,
       volume = {425},
       number = {3},
        pages = {2116-2127},
          doi = {10.1111/j.1365-2966.2012.21524.x},
archivePrefix = {arXiv},
       eprint = {1112.0312},
 primaryClass = {astro-ph.CO},
       adsurl = {https://ui.adsabs.harvard.edu/abs/2012MNRAS.425.2116C}
}

@ARTICLE{Lidman-2020,
       author = {{Lidman}, C. and {Tucker}, B.~E. and {Davis}, T.~M. and {Uddin}, S.~A. and {Asorey}, J. and {Bolejko}, K. and {Brout}, D. and {Calcino}, J. and {Carollo}, D. and {Carr}, A. and {Childress}, M. and {Hoormann}, J.~K. and {Foley}, R.~J. and {Galbany}, L. and {Glazebrook}, K. and {Hinton}, S.~R. and {Kessler}, R. and {Kim}, A.~G. and {King}, A. and {Kremin}, A. and {Kuehn}, K. and {Lagattuta}, D. and {Lewis}, G.~F. and {Macaulay}, E. and {Malik}, U. and {March}, M. and {Martini}, P. and {M{\"o}ller}, A. and {Mudd}, D. and {Nichol}, R.~C. and {Panther}, F. and {Parkinson}, D. and {Pursiainen}, M. and {Sako}, M. and {Swann}, E. and {Scalzo}, R. and {Scolnic}, D. and {Sharp}, R. and {Smith}, M. and {Sommer}, N.~E. and {Sullivan}, M. and {Webb}, S. and {Wiseman}, P. and {Yu}, Z. and {Yuan}, F. and {Zhang}, B. and {Abbott}, T.~M.~C. and {Aguena}, M. and {Allam}, S. and {Annis}, J. and {Avila}, S. and {Bertin}, E. and {Bhargava}, S. and {Brooks}, D. and {Carnero Rosell}, A. and {Carrasco Kind}, M. and {Carretero}, J. and {Castander}, F.~J. and {Costanzi}, M. and {da Costa}, L.~N. and {De Vicente}, J. and {Doel}, P. and {Eifler}, T.~F. and {Everett}, S. and {Fosalba}, P. and {Frieman}, J. and {Garc{\'\i}a-Bellido}, J. and {Gaztanaga}, E. and {Gruen}, D. and {Gruendl}, R.~A. and {Gschwend}, J. and {Gutierrez}, G. and {Hartley}, W.~G. and {Hollowood}, D.~L. and {Honscheid}, K. and {James}, D.~J. and {Kuropatkin}, N. and {Li}, T.~S. and {Lima}, M. and {Lin}, H. and {Maia}, M.~A.~G. and {Marshall}, J.~L. and {Melchior}, P. and {Menanteau}, F. and {Miquel}, R. and {Palmese}, A. and {Paz-Chinch{\'o}n}, F. and {Plazas}, A.~A. and {Roodman}, A. and {Rykoff}, E.~S. and {Sanchez}, E. and {Santiago}, B. and {Scarpine}, V. and {Schubnell}, M. and {Serrano}, S. and {Sevilla-Noarbe}, I. and {Suchyta}, E. and {Swanson}, M.~E.~C. and {Tarle}, G. and {Tucker}, D.~L. and {Varga}, T.~N. and {Walker}, A.~R. and {Wester}, W. and {Wilkinson}, R.~D. and {DES Collaboration}},
        title = "{OzDES multi-object fibre spectroscopy for the Dark Energy Survey: results and second data release}",
      journal = {\mnras},
         year = 2020,
        month = jul,
       volume = {496},
       number = {1},
        pages = {19-35},
          doi = {10.1093/mnras/staa1341},
archivePrefix = {arXiv},
       eprint = {2006.00449},
 primaryClass = {astro-ph.CO},
       adsurl = {https://ui.adsabs.harvard.edu/abs/2020MNRAS.496...19L}
}

@ARTICLE{Cardamone-2010,
       author = {{Cardamone}, Carolin N. and {van Dokkum}, Pieter G. and {Urry}, C. Megan and {Taniguchi}, Yoshi and {Gawiser}, Eric and {Brammer}, Gabriel and {Taylor}, Edward and {Damen}, Maaike and {Treister}, Ezequiel and {Cobb}, Bethany E. and {Bond}, Nicholas and {Schawinski}, Kevin and {Lira}, Paulina and {Murayama}, Takashi and {Saito}, Tomoki and {Sumikawa}, Kentaro},
        title = "{The Multiwavelength Survey by Yale-Chile (MUSYC): Deep Medium-band Optical Imaging and High-quality 32-band Photometric Redshifts in the ECDF-S}",
      journal = {\apjs},
         year = 2010,
        month = aug,
       volume = {189},
       number = {2},
        pages = {270-285},
          doi = {10.1088/0067-0049/189/2/270},
archivePrefix = {arXiv},
       eprint = {1008.2974},
 primaryClass = {astro-ph.CO},
       adsurl = {https://ui.adsabs.harvard.edu/abs/2010ApJS..189..270C}
}

@ARTICLE{Bacon-2023,
       author = {{Bacon}, Roland and {Brinchmann}, Jarle and {Conseil}, Simon and {Maseda}, Michael and {Nanayakkara}, Themiya and {Wendt}, Martin and {Bacher}, Raphael and {Mary}, David and {Weilbacher}, Peter M. and {Krajnovi{\'c}}, Davor and {Boogaard}, Leindert and {Bouch{\'e}}, Nicolas and {Contini}, Thierry and {Epinat}, Beno{\^\i}t and {Feltre}, Anna and {Guo}, Yucheng and {Herenz}, Christian and {Kollatschny}, Wolfram and {Kusakabe}, Haruka and {Leclercq}, Floriane and {Michel-Dansac}, L{\'e}o and {Pello}, Roser and {Richard}, Johan and {Roth}, Martin and {Salvignol}, Gregory and {Schaye}, Joop and {Steinmetz}, Matthias and {Tresse}, Laurence and {Urrutia}, Tanya and {Verhamme}, Anne and {Vitte}, Eloise and {Wisotzki}, Lutz and {Zoutendijk}, Sebastiaan L.},
        title = "{The MUSE Hubble Ultra Deep Field surveys: Data release II}",
      journal = {\aap},
         year = 2023,
        month = feb,
       volume = {670},
          eid = {A4},
        pages = {A4},
          doi = {10.1051/0004-6361/202244187},
archivePrefix = {arXiv},
       eprint = {2211.08493},
 primaryClass = {astro-ph.GA},
       adsurl = {https://ui.adsabs.harvard.edu/abs/2023A&A...670A...4B}
}

@ARTICLE{Talia-2023,
       author = {{Talia}, M. and {Schreiber}, C. and {Garilli}, B. and {Pentericci}, L. and {Pozzetti}, L. and {Zamorani}, G. and {Cullen}, F. and {Moresco}, M. and {Calabr{\`o}}, A. and {Castellano}, M. and {Fynbo}, J.~P.~U. and {Guaita}, L. and {Marchi}, F. and {Mascia}, S. and {McLure}, R. and {Mignoli}, M. and {Pompei}, E. and {Vanzella}, E. and {Bongiorno}, A. and {Vietri}, G. and {Amor{\'\i}n}, R.~O. and {Bolzonella}, M. and {Carnall}, A.~C. and {Cimatti}, A. and {Cresci}, G. and {Cristiani}, S. and {Cucciati}, O. and {Dunlop}, J.~S. and {Fontanot}, F. and {Franzetti}, P. and {Gargiulo}, A. and {Hamadouche}, M.~L. and {Hathi}, N.~P. and {Hibon}, P. and {Iovino}, A. and {Koekemoer}, A.~M. and {Mannucci}, F. and {McLeod}, D.~J. and {Saldana-Lopez}, A.},
        title = "{The VANDELS ESO public spectroscopic survey: The spectroscopic measurements catalogue}",
      journal = {\aap},
         year = 2023,
        month = oct,
       volume = {678},
          eid = {A25},
        pages = {A25},
          doi = {10.1051/0004-6361/202346293},
archivePrefix = {arXiv},
       eprint = {2309.14436},
 primaryClass = {astro-ph.GA},
       adsurl = {https://ui.adsabs.harvard.edu/abs/2023A&A...678A..25T}
}

@ARTICLE{Simon2019,
       author = {{Simon}, Joshua D.},
        title = "{The Faintest Dwarf Galaxies}",
      journal = {\araa},
         year = 2019,
        month = aug,
       volume = {57},
        pages = {375-415},
          doi = {10.1146/annurev-astro-091918-104453},
archivePrefix = {arXiv},
       eprint = {1901.05465},
 primaryClass = {astro-ph.GA},
       adsurl = {https://ui.adsabs.harvard.edu/abs/2019ARA&A..57..375S}
}

@ARTICLE{Kuijken-2019,
       author = {{Kuijken}, K. and {Heymans}, C. and {Dvornik}, A. and {Hildebrandt}, H. and {de Jong}, J.~T.~A. and {Wright}, A.~H. and {Erben}, T. and {Bilicki}, M. and {Giblin}, B. and {Shan}, H. -Y. and {Getman}, F. and {Grado}, A. and {Hoekstra}, H. and {Miller}, L. and {Napolitano}, N. and {Paolilo}, M. and {Radovich}, M. and {Schneider}, P. and {Sutherland}, W. and {Tewes}, M. and {Tortora}, C. and {Valentijn}, E.~A. and {Verdoes Kleijn}, G.~A.},
        title = "{The fourth data release of the Kilo-Degree Survey: ugri imaging and nine-band optical-IR photometry over 1000 square degrees}",
      journal = {\aap},
         year = 2019,
        month = may,
       volume = {625},
          eid = {A2},
        pages = {A2},
          doi = {10.1051/0004-6361/201834918},
archivePrefix = {arXiv},
       eprint = {1902.11265},
 primaryClass = {astro-ph.GA},
       adsurl = {https://ui.adsabs.harvard.edu/abs/2019A&A...625A...2K}
}
%



\begin{appendix}
\section{Random Forest performance with alternative photometric feature sets}
\label{app:rf_other_config}

In this section, we present a complementary suite of Random Forest experiments in which the classifier is trained and evaluated by excluding individual filters to quantify the robustness of our main results against changes in the available photometric information.

\subsection{Random Forest performance without the $u$ band}

Our first experiment consisted of removing the $u$-band from the reference input feature.
In this configuration, the sample increases to 15,207 sources (about 6\% more than the 14,360 objects with complete six-band coverage). As before, we allocate 70\% of the sample to the training set and the remaining 30\% to the validation set.
We repeated the hyperparameter optimization described in Sect.~\ref{sec:rf_tuning}, maximizing the F1 score on the validation sample.
Two distinct hyperparameter combinations achieved the best F1 score of 92.9\%, slightly below the value obtained with the reference feature set (93.3\%) but above the score obtained when removing the morphological parameter (90.1\%).
The two optimal configurations share the following parameters:
bootstrap = False, criterion = gini, max\_depth = 10, min\_samples\_leaf = 2, min\_samples\_split = 5, n\_estimators = 100,
while the only varying hyperparameter is max\_features (sqrt or log2).\par
\begin{figure}
    \centering
    \includegraphics[width=.9\linewidth]{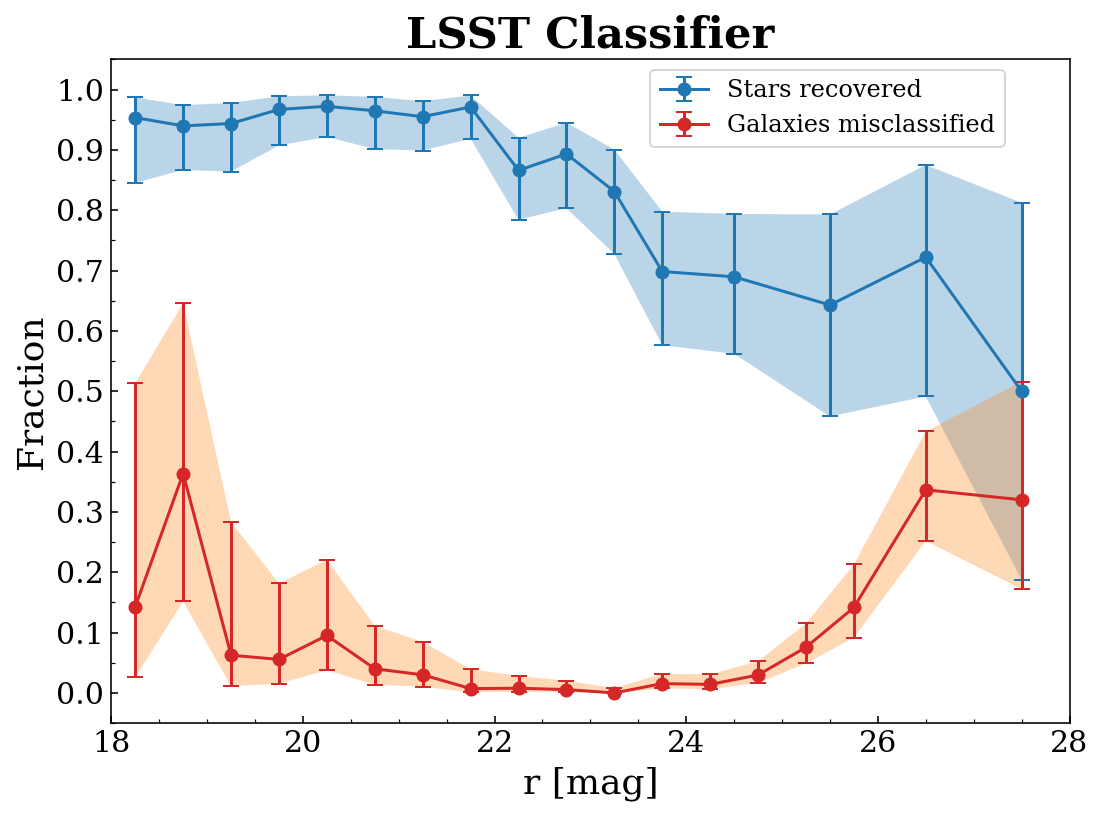}\\
    \includegraphics[width=.9\linewidth]{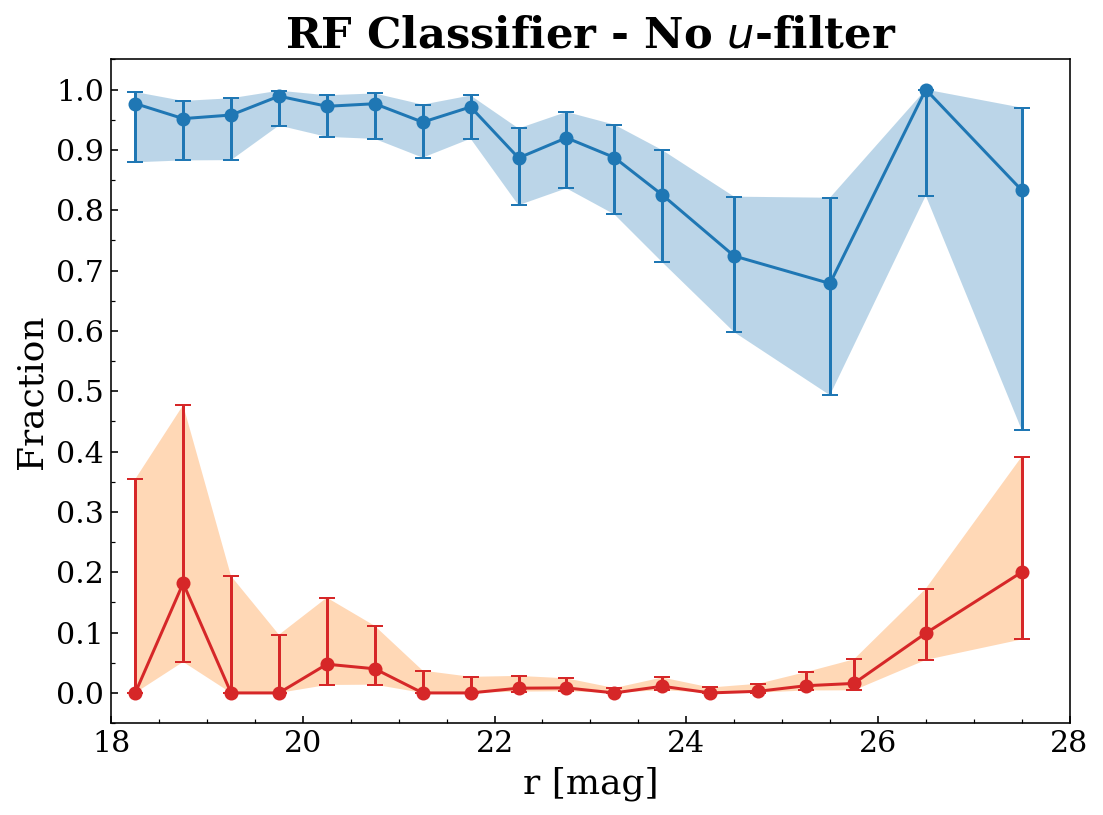}\\
    \caption{Same as Figure~\ref{fig:classification_fraction_result}, but computed on the validation sample for the experiment in which the $u$ band is excluded from the input feature set.}
    \label{fig:classification_fraction_result_nou}
\end{figure}

The complete set of performance metrics is reported in Table~\ref{tab:metrics}.
Overall, these metrics lie between those of the reference feature set and those obtained without {\sc refExtendedness}, suggesting that omitting the $u$-band is less disadvantageous than omitting the morphological information, at least when evaluated through global metrics alone.
However, Figure~\ref{fig:classification_fraction_result_nou} reveals a more sophisticated picture.
As in the leftmost panel of Fig.~\ref{fig:classification_fraction_result}, the upper panel illustrates the classification capability of {\sc refExtendedness} alone. Because of the slightly larger sample, this plot differs marginally from the one presented earlier, although the global trends remain unchanged. The morphological classifier has a good performance in identifying stars down to $r \lesssim 23.5$ mag, beyond which its discriminating power steadily declines, reaching a stellar recovery fraction of only $\sim 50\%$ in the faintest bins. For galaxies, {\sc refExtendedness} remains highly reliable down to $r \lesssim 25$ mag, after which the misclassification rate rapidly increases, consistent with the behaviour seen in Fig.~\ref{fig:classification_fraction_result}.
The application of the Random Forest improves upon the performance of the morphological classifier alone.
Nevertheless, in the absence of the $u$ band, the contamination from misclassified galaxies at faint magnitudes is higher than in any experiment involving the full set of LSST colors, whether or not {\sc refExtendedness} is included. In the last magnitude bin, the contamination from galaxies rises to $\sim 20\%$.\par
This result suggests that the $u$-band is essential for suppressing galaxy contamination in the faint regime where LSST will achieve its full scientific reach.
Although the global metrics might suggest that removing {\sc refExtendedness} is more disadvantageous than removing the $u$ band, the magnitude dependent behaviour reveals the opposite: for applications such as UFD searches, where faint-end purity is critical, ultraviolet information outperforms morphological information.
In other words, retaining the $u$ band is more important than retaining the {\sc refExtendedness} parameter for preserving classification quality at the faint photometric limits.
Finally, we note that in the faintest two magnitude bins, the stellar recovery fraction increases. This behaviour likely reflects the very small number of stars in these bins (46 stars for $r \geq 25$ mag and 21 stars for $r \geq 26$ mag), combined with the model’s tendency to classify sources in regions of strong color degeneracy as stars.
This effect enhances stellar recall but simultaneously inflates galaxy misclassification.
This experiment reinforce the conclusion that the $u$ band provides the most critical information for maintaining effective star–galaxy separation at faint magnitudes.

\subsection{Random Forest performance without the $y$ band}

In this experiment, we evaluate the performance of the Random Forest classifier when the $y$-band is removed from the reference feature set.
The resulting sample contains 14,612 sources. As in the previous tests, we performed a full hyperparameter optimization following the procedure described in Sect.~\ref{sec:rf_tuning}.
Six distinct hyperparameter configurations achieved the highest F1 score of 93.4\%, a value marginally higher than that obtained with the reference feature set (93.3\%).
All optimal configurations share the following parameters:
bootstrap = False, criterion = gini, min\_samples\_leaf = 2, min\_samples\_split = 10, n\_estimators = 300.
The only hyperparameters that vary across the best-performing models are max\_depth (None, 50, or 100) and max\_features (sqrt or log2).\par
The performance metrics reported in Table~\ref{tab:metrics} show that excluding the $y$ band results in comparable, slight better performance than removing the $u$ band. 
Figure~\ref{fig:classification_fraction_result_noy} provides additional insight into the behaviour of the classifier as a function of magnitude. 
Overall, the fraction of misclassified galaxies shows a slight improvement compared to the case where the $u$-filter was excluded from the feature set, except for the faintest bin, where the contamination level reaches $\sim 30\%$.
This experiment reinforces the conclusion that having access to the full six-band LSST photometry is more important for reliable star–galaxy classification than including the \Extend~parameter at the expense of one filter, when high stellar purity is required. In particular, losing even a single band degrades performance more severely than omitting the morphological parameter, augmenting the risk that compact clusters or group galaxies are misidentified as UFDs.

\begin{figure}
    \centering
    \includegraphics[width=.9\linewidth]{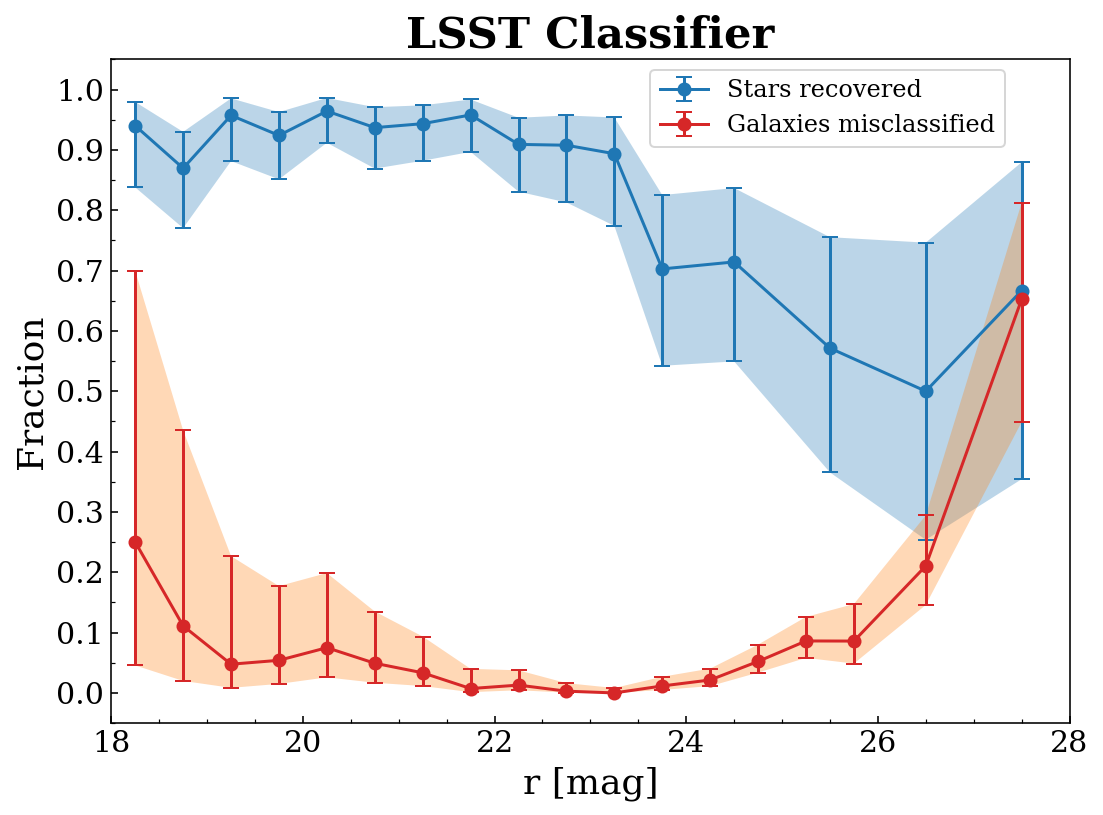}\\
    \includegraphics[width=.9\linewidth]{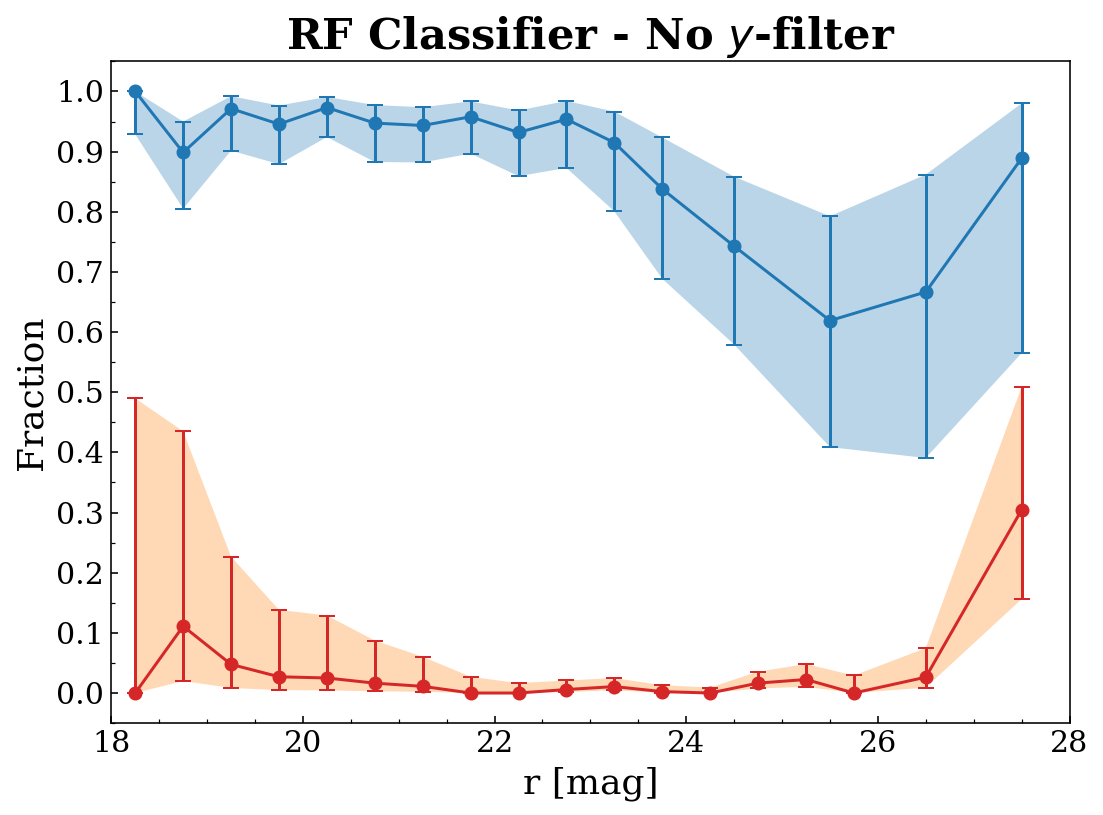}\\
    \caption{Same as Figure~\ref{fig:classification_fraction_result}, but computed on the validation sample for the experiment in which the $y$ band is excluded from the input feature set.}
    \label{fig:classification_fraction_result_noy}
\end{figure}

\subsection{Impact of photometric uncertainties in the absence of morphological features}
\label{app:photoerr_only}

In our last experiment, the classifier is trained using LSST multi-band photometry and the corresponding photometric uncertainties, while explicitly excluding the {\sc refExtendedness} parameter from the feature set. The goal of this test is to assess whether photometric uncertainties alone can compensate for the lack of morphological information, particularly in the faint regime where morphology becomes unreliable.\par
The hyperparameter optimization (see Sect.~\ref{sec:rf_tuning}) identified six distinct configurations that achieved the maximum F1 score of 90.1\%.
All optimal configurations share the same core set of hyperparameters, namely:
bootstrap = False, criterion = entropy, min\_samples\_leaf = 1, min\_samples\_split = 2, n\_estimators = 300.
The only hyperparameters that vary among these best-performing configurations are max\_depth (None, 50, or 100) and max\_features (sqrt or log2).\par
\begin{figure}
    \centering
    \includegraphics[width=.9\linewidth]{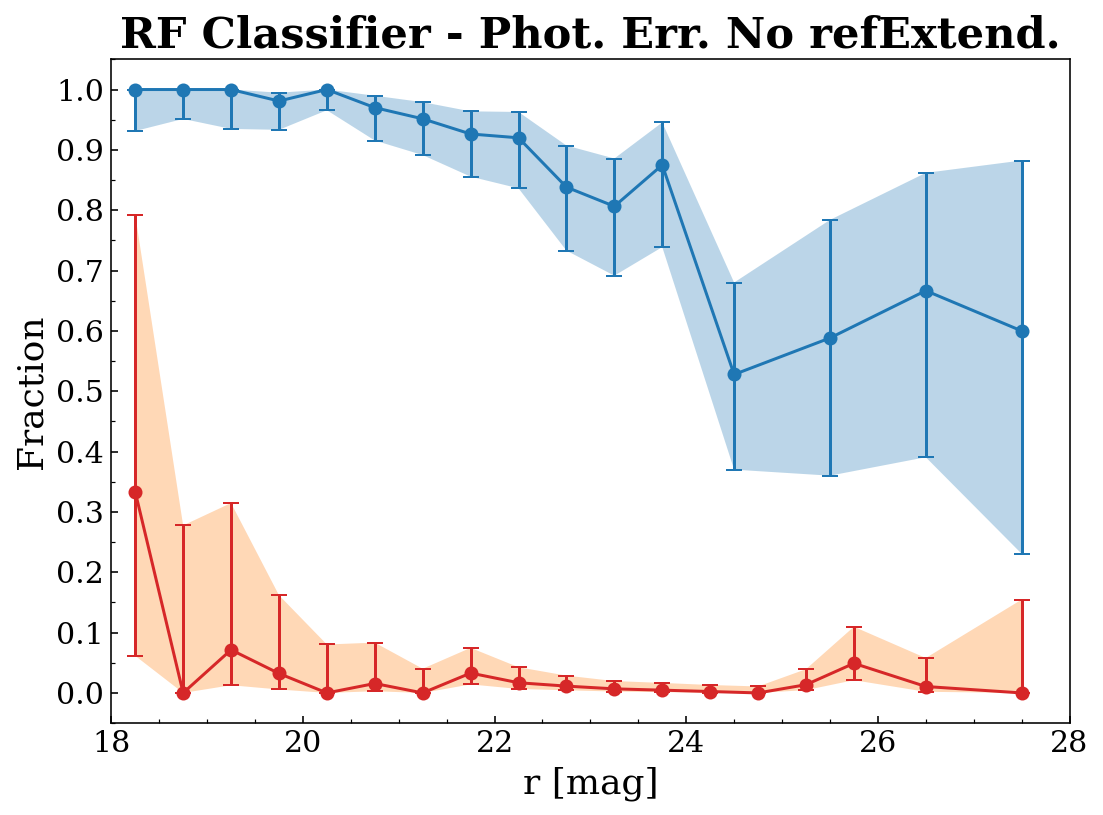}\\
    \caption{Same as Figure~\ref{fig:classification_fraction_result}, but computed on the validation sample for the experiment in which the \Extend~parameter is excluded and photometric uncertainties are included among the input features.}
    \label{fig:classification_fraction_result_noy}
\end{figure}
The performance metrics reported in Table~\ref{tab:metrics} show that, even in the absence of the morphological parameter, the inclusion of photometric uncertainties yields results only slightly inferior to those obtained with the reference input feature set.
For instance, the accuracy reaches 97.3\%, compared to 97.8\% for the reference configuration.
Notably, relative to the experiment in which {\sc refExtendedness} was excluded and photometric uncertainties were not included, the metrics associated with the stellar class improve substantially.
In particular, the stellar recall increases from 91.7\% to 92.9\%,  approaching the value of 94.3\% obtained with the reference feature set.\par
Figure~\ref{fig:classification_fraction_result_noy}\footnote{Note that we did not display the figure for the LSST internal classifier in this case as it would be identical to that displayed in the leftmost panel of Fig.~\ref{fig:classification_fraction_result}.} reveals that the fraction of misclassified galaxies even shows a slight improvement at the faintest magnitudes compared to the reference featuere set (left-center panel of Fig.~\ref{fig:classification_fraction_result}).
This behaviour indicates that the morphological parameter does not carry critical information for the correct classification of compact galaxies at faint magnitudes.\par
When comparing the two configurations that exclude the morphological parameter, the inclusion of photometric uncertainties leads to a marginally more stable recovery of stars at intermediate and faint magnitudes.
Although the differences are modest and largely consistent within the uncertainties, this trend suggests that photometric errors help the classifier to better accommodate stars that are scattered away from their intrinsic, narrow color–color loci.
In this context, the reduced stellar recall observed at faint magnitudes in the right-center panel of Fig.~\ref{fig:classification_fraction_result} appears to be driven primarily by increasing photometric scatter rather than by the absence of morphological information, which can be slightly mitigated by incorporating photometric uncertainties into the input features.\par
Finally, we note that the photometric precision and depth achieved in DP1 is lower than that expected for the final LSST survey (see Sect.~\ref{sec:intro}).
Consequently, future LSST data, benefiting from a deeper and more precise photometry, are likely to yield an even higher stellar recall at faint magnitudes, further mitigating the limitations observed in the present analysis.

\section{Comparison with XGBoost}
\label{app:xgboost}

{In this section, we describe an additional experiment using the XGBoost algorithm \citep{XGBoost-2016, XGBoost-2026}. 
For a direct comparison with the Random Forest, we restricted this test to the reference feature set (i.e. all LSST color combinations together with the {\sc refExtendedness} parameter). 
We optimized the hyperparameters following the same procedure described in Sect.~\ref{sec:rf_tuning}. 
The explored grid included: }
\begin{itemize}
    \item n\_estimators = 100, 300, 500, 1000;
    \item max\_depth = 3, 5, 7;
    \item min\_child\_weight = 1, 5;
    \item learning\_rate = 0.01, 0.05, 0.1;
    \item subsample = 0.8, 1.0;
    \item colsample\_bytree = 0.8, 1.0;
    \item gamma = 0, 0.1, 0.3.
\end{itemize}
{This corresponds to 864 distinct hyperparameter configurations. The best-performing model corresponds to the configuration:
colsample\_bytree = 1.0, gamma = 0, learning\_rate = 0.05, max\_depth = 7, min\_child\_weight = 1, n\_estimators = 300, subsample = 0.8.}\par
\begin{table}[]
\caption{Performance metrics for the XGBoost classifier on the validation sample for the reference feature set.}
    \tiny
    \begin{tabular}{l|l|c|c|c|c}
    \hline
    Features & Class & Precision & Recall & F1 score & N objects\\
     &  & \% & \% & \% & \\
      \hline\hline
      \multirow{2}{*}{Reference Set} & Galaxies & 98.1 & 98.7 & 98.4 & 3109\\
     & Stars & 96.6 & 95.0 & 95.8 & 1199\\
         \hline 
         \multicolumn{6}{c}{Accuracy = 97.7\%}\\
         \hline\hline
    \end{tabular}
    
    \label{tab:metrics_xgboost}
\end{table}
{The resulting performance metrics on the validation sample are reported in Table~\ref{tab:metrics_xgboost}. The overall accuracy is 97.7\%, namely 0.1\% lower than that obtained with Random Forest.
For galaxies, XGBoost yields a slightly higher precision (+0.3\%) but a marginally lower recall (-0.4\%). For stars, the recall increases by 0.7\%, but the precision decreases by 1.0\%, implying a slightly higher contamination of galaxies within the stellar sample. For both classes, the F1 score is lower by 0.1\% compared to Random Forest.}
\begin{figure}
    \centering
    \includegraphics[width=.9\linewidth]{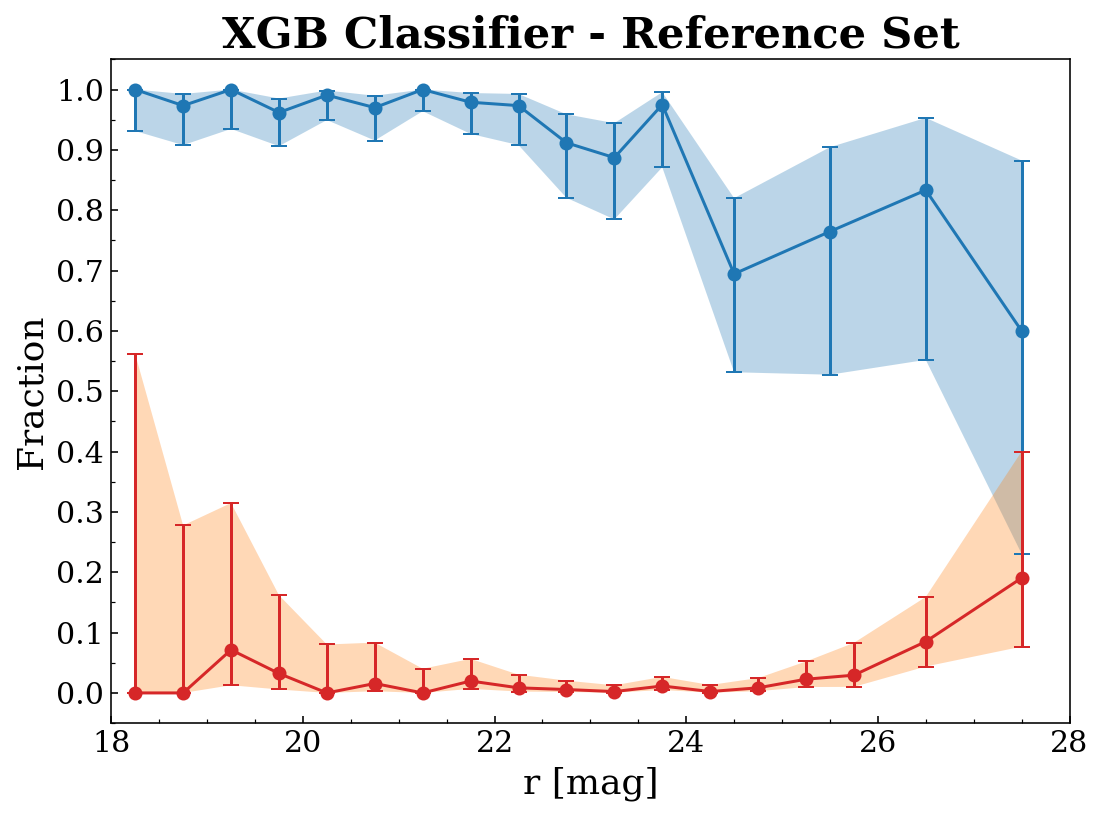}
        \caption{Same as Figure~\ref{fig:classification_fraction_result}, but computed with the XGBoost classifier.}
    \label{fig:classification_fraction_result_xgboost}
\end{figure}
{Figure~\ref{fig:classification_fraction_result_xgboost} shows the magnitude-dependent behavior. At faint magnitudes ($r \gtrsim 26$ mag), XGBoost exhibits a slightly earlier rise in galaxy contamination compared to Random Forest, reaching approximately 20\% in the last magnitude bin. The stellar recovery fraction also decreases to about 60\% in the faintest bin.}\par
{Overall, XGBoost delivers performance fully consistent with Random Forest, with marginally lower global metrics and slightly higher contamination at the faint end. Given our primary goal of minimizing galaxy contamination in deep stellar catalogs, we adopted Random Forest as the reference classifier throughout this work. 
We note, however, that XGBoost is significantly more computationally efficient: the hyperparameter search was approximately 45 times faster than for Random Forest. For larger datasets or more extensive hyperparameter grids, XGBoost may therefore represent a competitive alternative.}

\section{Comparison of predicted and true stellar locus}
\label{app:color_color_true_pred}

\begin{figure}
    \centering
    \includegraphics[width=.9\linewidth]{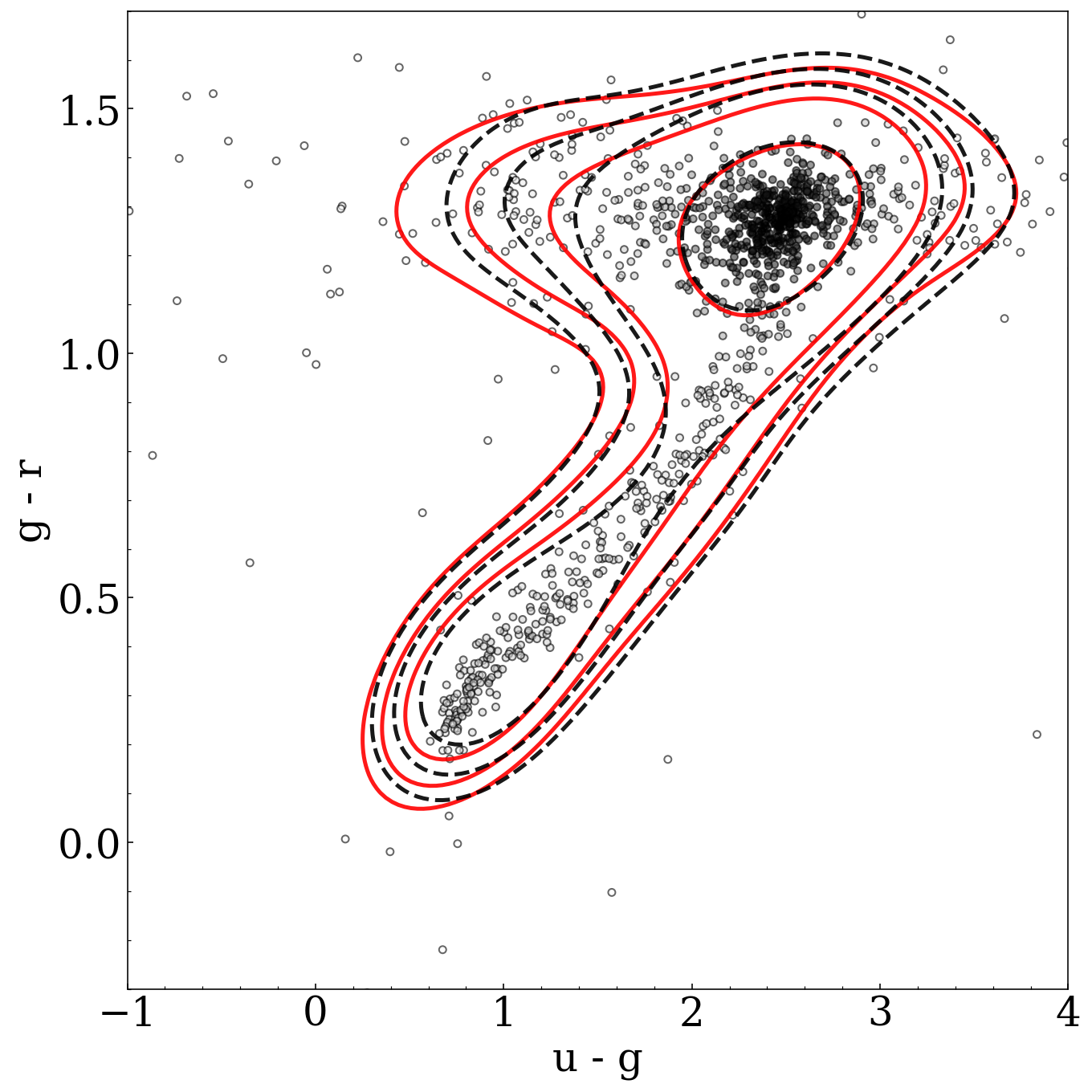}
        \caption{$(u-g)$ versus $(g-r)$ color-color diagram comparing objects classified as stars by the Random Forest model with bona-fide stars in our catalog. Colored points show the predicted stellar sample in the validation set. Black dashed contours represent the isodensity distribution of the predicted stars, while red solid contours correspond to the bona-fide stellar sample from the full catalogue.}
    \label{fig:stellar_locus_true_pred_stars}
\end{figure}

To further assess the robustness of the Random Forest, we compared the color–color distribution of objects classified as stars with the locus defined by our sample of bona-fide stars.
Figure~\ref{fig:stellar_locus_true_pred_stars} shows the $(u-g)$ versus $(g-r)$ diagram. The colored points represent objects classified as stars by the Random Forest model in the validation sample, while the black dashed contours indicate their isodensity distribution. The red solid contours correspond to the isodensity distribution of bona-fide stars from the full catalogue.
The predicted stellar sample closely follows the well-defined stellar locus over the entire color range. In particular, the high-density core of the classified stars overlaps remarkably well with that of the true stars, demonstrating that the classifier preserves the intrinsic structure of the stellar population in color-color space.
At lower densities, the agreement remains very good. Only a slight difference is visible in the bluest $(u-g)$ wing, where the predicted sample appears marginally less extended than the true one.

\end{appendix}

\end{document}